\documentclass[aps,twocolumn,showpacs,amssymb,eqsecnum]{revtex4}
\usepackage{epsfig}
\usepackage{graphicx}
\usepackage{dcolumn}
\usepackage{bm}

\begin{document}
\title
{The Kardar-Parisi-Zhang equation in the weak noise limit:\\
Pattern formation and upper critical dimension}
\author{Hans C. Fogedby}
\email{fogedby@phys.au.dk} \affiliation {Department of Physics and
Astronomy,
University of Aarhus, DK-8000, Aarhus C, Denmark\\
and\\
NORDITA, Blegdamsvej 17, DK-2100, Copenhagen {\O}, Denmark }
\begin{abstract}
We extend the previously developed weak noise scheme, applied to
the noisy Burgers equation in 1D, to the Kardar-Parisi-Zhang
equation for a growing interface in arbitrary dimensions. By means
of the Cole-Hopf transformation we show that the growth morphology
can be interpreted in terms of dynamically evolving textures of
localized growth modes with superimposed diffusive modes. In the
Cole-Hopf representation the growth modes are static solutions to
the diffusion equation and the nonlinear Schr\"odinger equation,
subsequently boosted to finite velocity by a Galilei
transformation. We discuss the dynamics of the pattern formation
and, briefly, the superimposed linear modes. Implementing the
stochastic interpretation we discuss kinetic transitions and in
particular the properties in the pair mode or dipole sector. We
find the Hurst exponent H=(3-d)/(4-d) for the random walk of
growth modes in the dipole sector. Finally, applying Derrick's
theorem based on constrained minimization we show that the upper
critical dimension is d=4 in the sense that growth modes cease to
exist above this dimension.
\end{abstract}
\maketitle
\section{\label{intro}Introduction}
The large majority of natural phenomena are characterized by being
out of equilibrium. This class includes turbulence in fluids,
interface and growth problems, chemical reactions, processes in
glasses and amorphous systems, biological processes, and even
aspects of economical and sociological structures.

As a consequence much of the focus of modern statistical physics,
soft condensed matter, and biophysics has shifted towards such
systems \cite{Nelson03,Chaikin95}. Drawing on the case of static
and dynamic critical phenomena in and close to equilibrium, where
scaling and the concept of universality have successfully served
to organize our understanding and to provide a variety of
calculational tools \cite{Binney92}, a similar strategy has been
advanced towards the much larger class of nonequilibrium phenomena
with the purpose of elucidating scaling properties and more
generally the morphology or pattern formation in a driven
nonequilibrium system \cite{Chaikin95,Cross94}.

There is a particular interest in the scaling properties and
general morphology of nonequilibrium models
\cite{Cross94,Barabasi95}. Here the Kardar-Parisi-Zhang (KPZ)
equation has played a prominent and paradigmatic role. The KPZ
equation describes aspects of the nonequilibrium kinetic growth of
a noise-driven interface and provides a simple continuum model of
an open driven nonlinear system exhibiting scaling and pattern
formation \cite{Kardar86,Medina89}.

The KPZ equation for the time evolution of the height field
$h({\bf r},t)$ has the form
\begin{eqnarray}
&&\frac{\partial h}{\partial t}=\nu{\bm\nabla}^2h
+\frac{\lambda}{2}{\bm\nabla} h\cdot{\bm\nabla} h-F+\eta,
\label{kpz}
\\
&&\langle\eta({\bf r},t)\eta({\bf
r}',t')\rangle=\Delta\delta^d({\bf r}-{\bf r}')\delta(t-t').
\label{noise}
\end{eqnarray}
Here the damping coefficient or surface tension $\nu$
characterizes the linear diffusion term, the parameter $\lambda$
controls the strength of the nonlinear growth term, $F$ is a
constant imposed drift term, and $\eta$ a locally correlated white
Gaussian noise modelling the stochastic nature of the drive or
environment; the noise correlations are characterized by the noise
strength $\Delta$
\cite{Krug97,Barabasi95,Krug92,Family90,Halpin95}.

In terms of the vector slope field
\begin{eqnarray}
{\bf u}={\bm\nabla}h, \label{vecslope}
\end{eqnarray}
the KPZ equation maps onto the Burgers equation driven by
conserved noise \cite{Forster76,Forster77,E00,Woyczynski98}
\begin{eqnarray}
\frac{\partial{\bf u}}{\partial t} =\nu{\bm\nabla}^2{\bf u}+
\lambda({\bf u}\cdot{\bm\nabla}){\bf u}+{\bm\nabla}\eta.
\label{bur}
\end{eqnarray}
In the deterministic case for $\eta=0$ the Burgers equation has
been used to study irrotational fluid motion and turbulence
\cite{Burgers29,Burgers74,Saffman68,Jackson90,Whitham74} and also
as a model for large scale structures in the universe
\cite{Zeldovitch72}.

In a series of papers we have analyzed the one dimensional noisy
Burgers equation for the slope field of a growing interface. In
Ref. \cite{Fogedby98a} we discussed as a prelude the noiseless
Burgers equation \cite{Burgers29,Burgers74} in terms of its
nonlinear soliton or shock wave excitations and performed a linear
stability analysis of the superimposed diffusive mode spectrum.
This analysis provided a heuristic picture of the damped transient
pattern formation. As a continuation of previous work on the
continuum limit of a spin representation of a solid-on-solid model
for a growing interface \cite{Fogedby95}, we applied in Ref.
\cite{Fogedby98b} the Martin-Siggia-Rose formalism \cite{Martin73}
in its path integral formulation
\cite{Baussch76,Janssen76,deDominicis78} to the noisy Burgers
equation \cite{Forster76,Forster77} and discussed in the weak
noise limit the growth morphology and scaling properties in terms
of nonlinear soliton or domain wall excitations with superimposed
linear diffusive modes. In Ref. \cite{Fogedby99a} we pursued a
canonical phase space approach based on the weak noise saddle
point approximation to the Martin-Siggia-Rose functional or,
alternatively, the Freidlin-Wentzel symplectic approach to the
Fokker-Planck equation \cite{Freidlin98,Graham89}. This method
provides a dynamical system theory point of view
\cite{Lichtenberg83,Ott93,Schuster89} to weak noise stochastic
processes and yields direct access to the probability
distributions for the noisy Burgers equation; brief accounts of
the above works has appeared in Refs.
\cite{Fogedby98c,Fogedby99b}. Further work on the scaling
properties and a numerical investigation of domain wall collisions
has appeared in in Refs.
\cite{Fogedby01a,Fogedby01b,Fogedby02a,Fogedby02b, Fogedby02c}. A
detailed summary and further developments have been given in an
extensive paper \cite{Fogedby03b}.

In the present work we address the KPZ equation for a growing
interface in arbitrary dimensions. Applying an extended form of
the canonical weak noise approach in order to incorporate
multiplicative noise and drawing from the insight gained by the
analysis of the 1D noisy Burgers equation, we identify the
localized growth modes for the KPZ equation. The growth modes are
spherically symmetric and are equivalent to the domain walls or
solitons identified in the 1D case. The growth modes propagate and
a dilute gas of modes constitute a dynamical network accounting
for the kinetic growth of the interface.

We also consider the issue of an upper critical dimension for the
KPZ equation. The KPZ equation lives at a critical point, conforms
to the dynamical scaling hypothesis
\cite{Barabasi95,Family86,Family90} and is characterized by the
scaling exponents $z$ and $\zeta$ \cite{Halpin95,Krug97}. Dynamic
renormalization group calculations yield $d=2$ as lower critical
dimension \cite{Forster77,Medina89}. In addition to the scaling
properties in the rough phase, characterized by a strong coupling
fixed point, a major open problem remains the existence of an
upper critical dimension \cite{Wiese98,Laessig97,Colaiori01a}. In
the present context we interpret the upper critical dimension as
the dimension beyond which the growth modes cease to exist. On the
basis of a numerical analysis and an exact argument based on
Derrick's theorem \cite{Derrick64} we propose that $d=4$ is the
upper critical dimension for the KPZ equation.

The paper is organized in the following way. To bring the reader
up to date we review in Sec.~\ref{kpzeq} the KPZ equation with
emphasis on the scaling properties. In Sec.~\ref{weak} we
summarize the weak noise approach including an extension to the
case of multiplicative noise in order to treat the KPZ equation in
the Cole-Hopf representation. In Sec.~\ref{kpzweak} we address the
KPZ equation and the associated noisy Burgers and Cole-Hopf
equations within the weak noise scheme and derive the fundamental
deterministic field equations governing the weak noise behavior.
In Sec.~\ref{field} we turn to the solutions of the field
equations. As a prelude we review the solutions in the 1D case and
then turn to the KPZ equation in its Cole-Hopf representation in
higher dimensions. In Sec.~\ref{pattern} we form a dynamical
network of growth modes accounting for the growth morphology of
the KPZ equation. We establish a field theory based on the picture
of the growth modes as charged monopoles. Finally, we discuss
briefly the superimposed linear mode spectrum. In
Sec.~\ref{scaling} we turn to the stochastic interpretation and
discuss kinetic transitions and in particular the anomalous
diffusion and scaling in the dipole sector. In Sec.~\ref{upper} we
discuss the issue of the upper critical dimension and present,
using Derrick's theorem based on constrained minimization, an
algebraic proof of the upper critical dimension being equal to
four. Sec.~\ref{sum} is devoted to a summary, a list of open
problems, and a conclusion. In appendix \ref{app} we consider the
application of the weak noise method to Brownian motion and the
overdamped oscillator. Aspects of the present work has appeared in
Ref. \cite{Fogedby05a}.
\section{\label{kpzeq} The KPZ equation}
The KPZ equation (\ref{kpz}) was proposed as a model for the
kinetic nonequilibrium growth of an interface driven by noise
\cite{Kardar86,Medina89,Barabasi95}. Although the equation only
describes limited aspects of true interface growth and, for
example, ignores surface diffusion \cite{Krug97}, the equation has
achieved an important and paradigmatic status in the theory of
nonequilibrium processes \cite{Halpin95}. In many regards the KPZ
equation serves as a prototype continuum model for nonequilibrium
processes in much the same way as the Ginzburg-Landau functional
in, for example, the context of critical phenomena
\cite{Binney92,Chaikin95}.
\subsection{General Properties}
From a structural point of view the KPZ equation (\ref{kpz}) has
the form of a noise-driven diffusion equation with a simple
nonlinear term added. Whereas the diffusion term $\nu{\bm\nabla}^2
h$ gives rise to a local flattening or relaxation of the
interface, corresponding to a surface tension, the crucial
nonlinear term $(\lambda/2){\bm\nabla}h\cdot{\bm\nabla}h$ accounts
for the lateral growth of the interface \cite{Barabasi95}. In that
sense the KPZ equation is a genuine kinetic equation describing a
nonequilibrium process in the sense that the drift term
$\nu{\bm\nabla}^2h+(\lambda/2){\bm\nabla}h\cdot{\bm\nabla}h-F$
cannot be derived from an effective free energy. The noise drives
the height field into a stationary state whose distribution is not
known in detail except in 1D, where it is independent of $\lambda$
and given by \cite{Huse85,Halpin95}
\begin{eqnarray}
P_0(h)\propto\exp\left[-\frac{\nu}{\Delta}\int
dx~({\bm\nabla}h)^2\right].\label{stakpz}
\end{eqnarray}
In the linear case for $\lambda=0$ the KPZ equation (\ref{kpz})
reduces to the Edwards-Wilkinson equation (EW) \cite{Edwards82}
\begin{eqnarray}
\frac{\partial h}{\partial t}=-\frac{1}{2}\frac{\delta{\cal
F}}{\delta h}-F+\eta~,~~ {\cal F}=\nu\int d^dx~({\bm\nabla}h)^2,
\label{ew}
\end{eqnarray}
which in a comoving frame with velocity $-F$ describes an
interface in thermal equilibrium at temperature $T=\Delta$ with
stationary Boltzmann distribution given by Eq. (\ref{stakpz}). In
the EW case we also have access to the time-dependent
distribution. Expanding the height field on a plane wave basis,
$h({\bf r})=\int d^dx~h_{\bf k}\exp(i{\bf kr})$ and introducing
the diffusive mode frequency $\omega_{\bf k}=\nu{\bf k}^2$ we
obtain for the transition probability from an initial profile
$h_{\bf k}^0$ to a final profile $h_{\bf k}$ in time $T$
\cite{Krug97}
\begin{eqnarray}
&&P(h_{k},T)\propto
\nonumber
\\
&&\left[-\frac{\nu}{\Delta}\int\frac{d^dk}{(2\pi)^d}~k^2
\frac{|h_{\bf k}-h_{\bf k}^0\exp(-\omega_{\bf k}T)|^2}
{1-\exp(-2\omega_{\bf k}T)} \right],~~ \label{disew}
\end{eqnarray}
and, for example, the height correlation function
\begin{eqnarray}
\langle hh\rangle({\bf k},\omega)=\frac{\Delta}{\omega^2+(\nu
k^2)^2}, \label{corew}
\end{eqnarray}
with a Lorentzian line shape controlled by the diffusive poles at
$\omega=\pm i\nu k^2$; see also Ref. \cite{Majaniemi96}.

Averaging the KPZ equation in a state at times where transients
have died out, we obtain
\begin{eqnarray}
\frac{d\langle
h\rangle}{dt}=(\lambda/2)\langle{\bm\nabla}h\cdot{\bm\nabla}h\rangle-F,
\label{avkpz}
\end{eqnarray}
showing that the nonlinear term gives rise to nonequilibrium
growth determined by the magnitude of
$\langle{\bm\nabla}h\cdot{\bm\nabla}h\rangle$; note that choosing
$F$ to balance the nonlinear term, i.e.,
$F=(\lambda/2)\langle{\bm\nabla}h\cdot{\bm\nabla}h\rangle$, is
equivalent to choosing a comoving frame in which $d\langle
h\rangle/dt=0$.

In addition to being invariant under time and space translations,
the KPZ equation (\ref{kpz}) is also invariant subject to the
nonlinear Galilei transformation:
\begin{eqnarray}
&&{\bf r}\rightarrow {\bf r}-\lambda{\bf u}^0t, \label{gal1}
\\
&&h\rightarrow h+{\bf u}^0\cdot{\bf r},\label{gal2}
\\
&&F\rightarrow F+(\lambda/2){\bf u}^0\cdot{\bf u}^0.\label{gal3}
\end{eqnarray}
Hence, the transformation to a moving frame with velocity
$\lambda{\bf u}^0$ is absorbed by adding a constant slope term
${\bf u}^0\cdot{\bf r}$ to the height field $h$ and shifting the
constant drift term by $(\lambda/2){\bf u}_0^2$. Note that the
invariance is associated with the nonlinear parameter $\lambda$
which enters as a structural constant in the Galilei group
transformation; in the EW case this invariance is absent.

\begin{figure}
\includegraphics[width=1.0\hsize]
{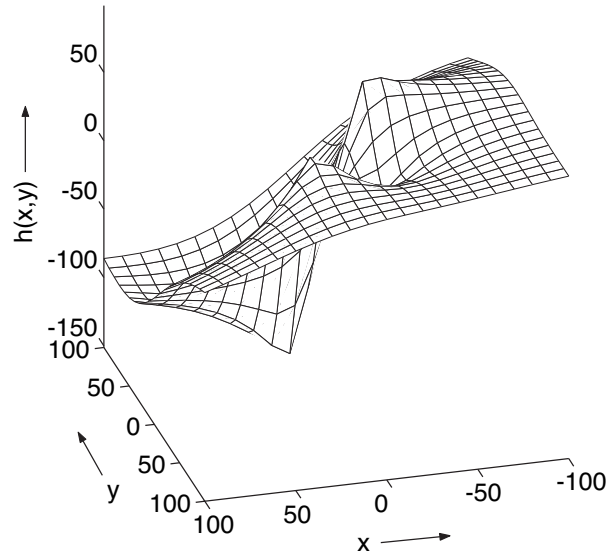} \caption{We depict a growing interface in 2D (arbitrary
units).} \label{fig1}
\end{figure}

The KPZ equation is characterized by the parameters $\nu$,
$\lambda$, $F$, and $\Delta$ of dimension $[\nu]=L^2/T$,
$[\lambda]=L/T$, $[F]=L/T$, and $[\Delta]=L^{d+2}/T$. By
transforming to a comoving frame $h\rightarrow h-Ft$ we can absorb
the drift $F$ for a given growth morphology and from the remaining
dimensionfull parameters form the dimensionless parameter
$\Delta\lambda^d/\nu^{d+1}$. Consequently, the weak noise limit
$\Delta\rightarrow 0$ is equivalent to the weak coupling limit
$\lambda\rightarrow 0$ (or the limit of large damping
$\nu\rightarrow\infty$). In Fig.~\ref{fig1} we have in 2D depicted
a growing interface.
%
\subsection{Burgers equation and Cole-Hopf equation}
The paradigmatic importance of the KPZ equation stems from the
fact that it in addition to describing nonequilibrium surface
growth also is associated with the theory of turbulence via its
equivalence to the noisy Burgers equation (\ref{bur}). Moreover,
by means of the nonlinear Cole-Hopf tranformation
\cite{Cole51,Hopf50,Medina89}
\begin{eqnarray}
w=\exp\left[\frac{\lambda}{2\nu}h\right], \label{ch}
\end{eqnarray}
the KPZ equation takes the form of a linear diffusion equation
driven by multiplicative noise, denoted here the Cole-Hopf
equation,
\begin{eqnarray}
\frac{\partial w}{\partial t} = \nu{\bm\nabla}^2w
-\frac{\lambda}{2\nu}w F+\frac{\lambda}{2\nu}w\eta. \label{che}
\end{eqnarray}
In the absence of noise Eq. (\ref{che}) becomes the linear
diffusion equation and is readily analyzed, thus allowing a
discussion of the KPZ and Burgers equations in the deterministic
case \cite{Medina89,Woyczynski98}. In the noisy case a path
integral representation of the solution of Eq. (\ref{che}) maps
the Cole-Hopf equation and thus the KPZ equation onto a model of a
directed polymer with line tension $1/4\nu$ in a quenched random
potential $(\lambda/2\nu)\eta({\bf r},t)$. The disordered directed
polymer model is a toy model within the spin glass literature and
has been analyzed by means of the replica method and Bethe ansatz
techniques \cite{Kardar87a,Halpin95}.

The Galilean invariance of the KPZ equation is for the Burgers
equation (\ref{bur}) and the Cole-Hopf equation (\ref{che})
supplemented by the transformations
\begin{eqnarray}
&&{\bf u}\rightarrow{\bf u}+{\bf u}^0,\label{gal4}
\\
&&w\rightarrow w\exp\left[\left(\lambda/2\nu\right){\bf
u}^0\cdot{\bf r}\right], \label{gal5}
\end{eqnarray}
i.e., the slope field is shifted by ${\bf u}^0$ and the Cole-Hopf
field by the multiplicative factor $\exp[(\lambda/2\nu){\bf
u}^0\cdot{\bf r}]$.
\subsection{Scaling properties}
The KPZ equation conforms to the dynamical scaling hypothesis with
long time-large distance height correlations
\cite{Barabasi95,Family85}
\begin{eqnarray}
\langle h h\rangle({\bf r},t) = r^{2\zeta}\Phi(t/r^z),
\label{corkpz}
\end{eqnarray}
characterized by the roughness exponent $\zeta$, dynamical
exponent $z$, and scaling function $\Phi$. Consequently, most
theoretical efforts have addressed the scaling issues. Based on i)
perturbative dynamic renormalization group (DRG) calculations in
combination with the scaling law
\begin{eqnarray}
\zeta+z=2,\label{scal}
\end{eqnarray}
following from Galilean invariance, and the known stationary
height distribution in 1D given by Eq. (\ref{stakpz})
\cite{Medina89,Huse85,Wiese97}, ii) the mapping of the KPZ
equation onto directed polymers in a quenched environment and
ensuing replica calculations \cite{Halpin95,Kardar87a}, iii) mode
coupling calculations
\cite{Schwartz92,Bouchaud93,Doherty94,Moore95}, iv) operator
expansion methods \cite{Laessig98a}, and v) numerical calculations
\cite{Marinari00,Marinari02,Newman97b,Beccaria94}, the following
picture has emerged.

\begin{figure}
\includegraphics[width=1.0\hsize]
{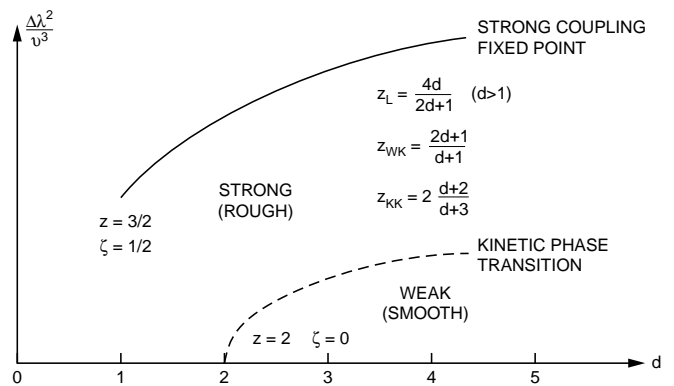}\caption{ We summarize the scaling properties of the KPZ
equation in a plot of the renormalized coupling constant (the
fixed point value) $\Delta\lambda^2/\nu^3$ versus the dimension
$d$ of the system. Below $d=2$ the scaling properties are
determined by the strong coupling fixed. Above $d=2$ the system
exhibits a kinetic phase transition from a weak coupling phase to
a strong coupling phase. We have also indicated three conjectures
for the exponent $z$: The operator expansion conjecture $z_L$ by
L\"{a}ssig \cite{Laessig98a} and the conjectures based on
numerics, $z_{WK}$ by Wolf and Kert\'{e}sz \cite{Wolf97} and
$z_{KK}$ by Kim and Kosterlitz \cite{Kim89}. } \label{fig2}
\end{figure}

In 1D the interface is rough and  characterized by a
perturbatively inaccessible strong coupling fixed point with
scaling exponents $z=3/2$ and $\zeta=1/2$, following from the
stationary distribution in combination with the scaling law. Above
the lower critical dimension $d=2$ a DRG calculation in
$\epsilon=d-2$ predicts a kinetic transition line between a smooth
phase characterized by a weak coupling fixed point at $\lambda=0$
with exponents $z=2$ and $\zeta=(2-d)/2$ and a rough phase
characterized by a poorly understood strong coupling fixed point.
On the transition line $z=2$ and $\zeta=0$
\cite{Laessig95,Wiese98}. Based on numerics the following
expressions for $z$ have been proposed:
$z_{\text{KK}}=2(d+2)/(d+3)$ \cite{Kim89} and
$z_{\text{WK}}=(2d+1)/(d+1)$ \cite{Wolf97}. Both $z_{\text{KK}}$
and $z_{\text{WK}}$ agree with $z=3/2$ in $d=1$; for
$d\rightarrow\infty$ we have $z\rightarrow 2$, corresponding to a
smooth phase at an infinite upper critical dimension. An operator
expansion method \cite{Laessig98a} yields $z_{\text{L}}=4d/(2d+1)$
for $d>1$ and $z\rightarrow 2$ for $d\rightarrow\infty$. The
scaling properties of the KPZ equation is summarized in
Fig.~\ref{fig2} in a plot of the renormalized coupling strength
$g=\lambda^2\Delta/\nu^3$ versus spatial dimension.

Whereas $d=2$ constitutes the lower critical dimension permitting
a loop expansion in powers of $\epsilon$, the issue of an upper
critical dimension has been much debated. Mode coupling approaches
yield an upper critical dimension $d=4$ with a possible glassy
behavior above 4
\cite{Moore95,Bhattacharjee98,Colaiori01a,Colaiori01b}. Loop
expansion to all orders in $\epsilon$ supported by an exact
evaluation of the beta-function in a Callen-Symanzik
renormalization group scheme associates the upper critical
dimension with a mathematical singularity
\cite{Laessig97,Wiese98}.

Two outstanding issues thus remain as regard the scaling
properties of the KPZ equation: The upper critical dimension and
the properties of the strong coupling fixed point above $d=1$,
i.e., the scaling exponents and scaling function. There is,
moreover, the more general question of the deeper mechanism behind
the stochastic growth morphology. In that respect the KPZ equation
shares its strong coupling features with another notable problem
in theoretical physics: Turbulence.

Since the DRG in its perturbative form as an expansion about the
linear theory fails to yield insight into the strong coupling
features there clearly is a need for alternative systematic
methods. Here we should like to emphasize that both the mode
coupling approaches
\cite{Moore95,Doherty94,Colaiori01a,Colaiori01b,
Bouchaud93,Bhattacharjee98} and operator expansion methods
\cite{Laessig98a} do not qualify as being systematic. The mode
coupling approach is based on a truncation procedure ignoring
vertex corrections and thus violating Galilean invariance which is
essential in delimiting the KPZ universality class. The operator
product expansion imposes an ad hoc mathematically motivated
operator structure.
\section{\label{weak} The weak noise method}
The weak noise method is based on an asymptotic weak noise
approximation to a general Langevin equation driven by Gaussian
white noise. The method dates back to Onsager
\cite{Onsager53,Machlup53} and has since reappeared as the
Freidlin-Wentzel theory of large deviations
\cite{Freidlin98,Graham89} and as the weak noise saddle point
approximation to the functional Martin-Siggia-Rose scheme
\cite{Martin73,Baussch76,Janssen76,deDominicis78}. The method
known also as the eikonal approximation has, moreover, been used
in the context of thermally activated escape
\cite{Graham84,Graham84a,Graham85,Graham89,Dykman90,
Bray89,Maier96,Maier01,Aurell02}. The weak noise or canonical
phase space method has been discussed in Ref. \cite{Fogedby99a}.
In the present context we need a generalization of the method in
order to accommodate multiplicative noise and therefore review and
extend the method below.
\subsection{General properties}
The point of departure is the generic Langevin equations for a set
of stochastic variables $w_n$, $n=1,\cdots N$, driven by white
Gaussian noise \cite{Stratonovich63}
\begin{eqnarray}
&&\frac{dw_p}{dt}=-\frac{1}{2}F_p-\frac{\Delta}{2}G_{mn}\nabla_mG_{pn}
+G_{pn}\eta_n, \label{lan}
\\
&&\langle\eta_n\eta_m\rangle(t)=\Delta\delta_{nm}\delta(t),
\label{wnoise}
\end{eqnarray}
where $F_n(w_q)$ is the drift, $G_{nm}(w_q)$ is accounting for
multiplicative noise, $\nabla_n=\partial/\partial w_n$, and
$\Delta$ is the explicit noise strength; sums are performed over
repeated indices. In the case of multiplicative noise with $G$
depending on $w$ this is the Stratonovich formulation. In the Ito
formulation the compensating drift term $G\nabla G$ is absent
corresponding to the application of the Ito differentiation rules
\cite{Risken89,Reimann02}. Note that in the case of i) additive
noise, i.e., $G$ independent of $w$ or ii) to leading order in
$\Delta$ the Ito and Stranovich formulations are equivalent. The
generic Langevin equation (\ref{lan}) driven by white noise
encompasses with appropriate choices of $F$ and $G$ all white
noise driven continuous Markov processes.

On the deterministic level the corresponding Fokker-Planck
equation for the probability distribution $P(w_n,t)$ has the form
\cite{Stratonovich63}
\begin{eqnarray}
\frac{\partial P}{\partial
t}=\frac{1}{2}\nabla_n[F_n+\Delta\nabla_mK_{mn}]P, \label{fp}
\end{eqnarray}
where the symmetrical noise matrix is given by
\begin{eqnarray}
K_{pm}(w_q)=G_{pn}(w_q)G_{mn}(w_q). \label{nm}
\end{eqnarray}

In complete analogy with the WKB approach in quantum mechanics
\cite{Landau59c}, $\Psi\propto\exp[iS/\hbar]$, relating the wave
function $\Psi$ to the classical action $S$ evaluated on the basis
of the classical Hamilton equations of motion, it is useful to
capture weak noise effects and relate the stochastic problem to a
scheme based on classical equations of motion by means of a weak
noise WKB approximation to the Fokker-Planck equation (\ref{fp}).
Thus introducing the Wentzel-Kramers-Brillouin (WKB) ansatz
\begin{eqnarray}
P(w_n,t)\propto\left[-\frac{S(w_n,t)}{\Delta}\right], \label{wkb}
\end{eqnarray}
we obtain to leading order in the noise strength $\Delta$ the
Hamilton-Jacobi equation \cite{Landau59c,Goldstein80}
\begin{eqnarray}
\frac{\partial}{\partial t} + H=0~,~~ p_n=\nabla_n S, \label{canp}
\end{eqnarray}
with Hamiltonian
\begin{eqnarray}
H=-\frac{1}{2}p_nF_n+\frac{1}{2}K_{nm}p_np_m. \label{ham}
\end{eqnarray}
The Hamiltonian equations of motion follow from $dw_n/dt=\partial
H/\partial p_n$ and $dp_n/dt=-\partial H/\partial w_n$,
\begin{eqnarray}
&&\frac{dw_n}{dt}=-\frac{1}{2}p_nF_n+K_{nm} p_m, \label{eq1}
\\
&&\frac{dp_n}{dt}=
\frac{1}{2}p_m\nabla_nF_m-\frac{1}{2}p_mp_q\nabla_nK_{mq},
\label{eq2}
\end{eqnarray}
determining classical orbits on the energy surfaces given by $H$
in a classical phase space $(\{w_n\},\{p_n\})$. Finally, the
action $S$ is given by
\begin{eqnarray}
S(w_n,T)=\int^{w_n,T}_{w_n^{\text{i}},0}dt~p_n\frac{dw_n}{dt} -
HT, \label{act}
\end{eqnarray}
yielding according to Eq. (\ref{wkb}) the transition probability.
By means of the equation of motion (\ref{eq1}) the action can be
reduced to the form
\begin{eqnarray}
S(w_n,T)=\frac{1}{2}\int^{w_n,T}_{w_n^{\text{i}},0}dt
~K_{nm}p_np_m. \label{act2}
\end{eqnarray}

The weak noise scheme bears the same relationship to stochastic
fluctuations as the WKB approximation in quantum mechanics,
associating the phase of the wave function with the action of the
corresponding classical orbit \cite{Landau59c}. In addition to
providing a classical orbit picture of stochastic fluctuations and
thus allowing the use of dynamical system theory
\cite{Arnold89,Jackson90,Lichtenberg83,Reichl87,Scott99,Schuster89},
the method also yields the Arrhenius factor
$P\propto\exp(-S/\Delta)$ for a kinetic transition from
$w_n^{\text{i}}$ to $w_n$ during the transition time $T$. Here the
action $S$ serves as the weight in the same manner as the energy
$E$ in the Boltzmann factor $P\propto\exp(-\beta E)$, $\beta =
1/kT$, for equilibrium processes.

In the weak noise scheme the stochastic Langevin equation
(\ref{lan}) is replaced by the deterministic equation of motion
(\ref{eq1}) for $w_n$ which together with the equation of motion
(\ref{eq2}) for the canonically conjugate noise variable $p_n$
determine orbits lying on the constant energy manifolds
$H=\text{const.}$ in a canonical phase space spanned by $w_n$ and
$p_n$. Determining a specific orbit from $w_n^{\text{i}}$ to $w_n$
in time $T$ by solving the equations of motion (\ref{eq1}) and
(\ref{eq2}) with $p_n$ as a slaved variable, evaluating the action
(\ref{act}), then yields the (unnormalized) transition probability
in Eq. (\ref{wkb}).

A stationary distribution of the Fokker-Planck equation (\ref{fp})
is given by
\begin{eqnarray}
P_0(w_n)=\lim_{T\rightarrow\infty}P(w_n,T), \label{stat}
\end{eqnarray}
satisfying $(1/2)\nabla_n[F_n+\Delta\nabla_mK_{mn}]P=0$. Within
the dynamical system theory framework to leading asymptotic order
in $\Delta$ this implies that $S(w_n,T)\rightarrow S_0(w_n)$ for
fixed final configuration $w_n$, i.e.,
$P_0\propto\exp[-S_0/\Delta]$. To attain the stationary limit an
orbit from an initial configuration $w_n^{\text{i}}$ to a final
configuration $w_n$ traversed in time $T\rightarrow\infty$ must
(i) pass through a saddle point and (ii) in the limit lie on a
zero-energy manifold $H=0$. The zero-energy manifold is in general
composed of two submanifolds intersecting at the saddle point. The
transient or noiseless submanifold $p=0$, yielding $H=0$ and
consistent with the equations of motion, corresponds to the
transient motion determined by the noiseless equation of motion
$dw_n/dt=-(1/2)p_nF_n$. The stationary or noisy submanifold
determined by the orthogonality condition $p_n[K_{nm}p_m -
F_n]=0$, corresponds to the stationary motion. The loss of memory
and Markovian behavior result from the infinite waiting time at
the saddle point; for details see e.g. Ref. \cite{Fogedby99a}.

\begin{figure}
\includegraphics[width=1.0\hsize]
{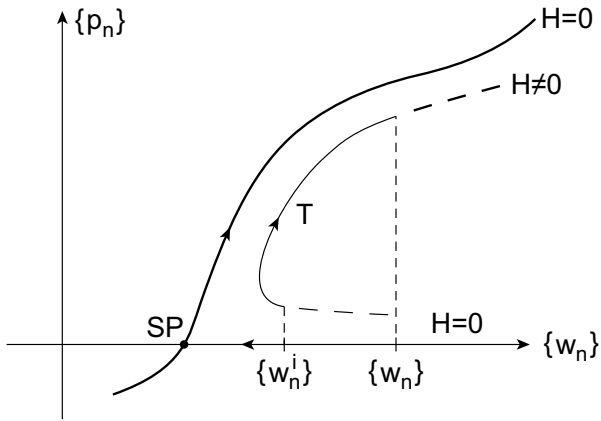}\caption{ The $(\{w_n\},\{p_n\})$ phase space with a
finite-time orbit from $\{w_n^i\}$ to $\{w_n\}$ in time $T$ on a
$H\neq 0$ energy manifold, yielding the transition probability,
and an infinite-time stationary orbit to $\{w_n\}$ passing through
the saddle point SP and lying on the $H=0$ energy manifold.}
\label{fig3}
\end{figure}

We, moreover, note that the weak noise scheme has a symplectic
structure with Poisson bracket algebra
\cite{Landau59b,Goldstein80}
\begin{eqnarray}
\{w_n,p_m\}=\delta_{nm},~~\{w_n,w_m\}=\{p_n,p_m\}=0, \label{b}
\end{eqnarray}
and Hamiltonian equations of motion
\begin{eqnarray}
\frac{dw_n}{dt}=\{w_n,H\}~,~~\frac{dp_n}{dt}=\{p_n,H\}. \label{h}
\end{eqnarray}
In Fig.~\ref{fig3} we have depicted the generic phase space
structure in the case where the system possesses a stationary
state.

The purpose of the weak noise scheme is twofold. On the one hand,
it provides an alternative way of discussing stochastic phenomena
in terms of an equivalent deterministic scheme based on canonical
equations of motion, orbits in phase space, and the ensuing
dynamical system theory concepts. The scheme, on the other hand,
also provides a calculational tool in determining the transition
probabilities and ensuing correlations. It is also important to
keep in mind that although the starting point is a weak noise
approximation the scheme, like WKB in quantum mechanics, is
nonperturbative and is thus capable of accounting at least
qualitatively for strong noise effects. This point of view was
stressed by Coleman in the context of quantum field theories
\cite{Coleman77}.

This completes our general discussion of the weak noise approach.
As an illustration we consider in appendix \ref{app} the
application of the method to two systems with a single degree of
freedom: (i) random walk and (ii) the overdamped oscillator. See
also an application to a nonlinear finite-time-singularity model
in Refs. \cite{Fogedby02d,Fogedby03c} and to an extended system,
the noise-driven Ginzburg-Landau model, in Refs
\cite{Fogedby03a,Fogedby04b}.

\section{\label{kpzweak} The KPZ equation for weak noise}
Here we apply the weak noise method to the KPZ equation, the
corresponding noisy Burgers equation and Cole-Hopf equation.
\subsection{The KPZ case}
Adapting the KPZ equation (\ref{kpz}) to the weak noise scheme we
make the assignment: $w_n(t)\rightarrow h({\bf
r},t),p_n(t)\rightarrow \tilde p({\bf r},t),K_{nm}\rightarrow
\delta^d ({\bf r}-{\bf r}'),F_n\rightarrow
-2\left[\nu{\bm\nabla}^2 h+(\lambda/2){\bm\nabla}
h\cdot{\bm\nabla} h -F\right]$. Here the index $n$ becomes the
spatial coordinate ${\bf r}$. We thus obtain the KPZ Hamiltonian
density
\begin{eqnarray}
{\cal H}^{\text{KPZ}}=\tilde
p\left[\nu{\bm\nabla}^2h+\frac{\lambda}{2}{\bm\nabla}
h\cdot{\bm\nabla} h -F+\frac{1}{2}\tilde p\right], \label{kpzham}
\end{eqnarray}
and the canonical field equations
\begin{eqnarray}
&&\frac{\partial h}{\partial t}=\nu{\bm\nabla}^2
h+\frac{\lambda}{2}{\bm\nabla} h\cdot{\bm\nabla} h-F+\tilde p,
\label{kpzeq1}
\\
&&\frac{\partial\tilde p}{\partial t}=-\nu{\bm\nabla}^2\tilde
p+\lambda{\bm\nabla}\tilde p\cdot{\bm\nabla} h+\lambda\tilde
p\nabla^2 h. \label{kpzeq2}
\end{eqnarray}
The reduced action and transition probability are given by
\begin{eqnarray}
&&S^{\text{KPZ}}(h,T)=\frac{1}{2}\int^{h,T} d^dxdt~\tilde p({\bf
r},t)^2, \label{kpzact}
\\
&&P^{\text{KPZ}}(h,T)\propto
\exp\left[-\frac{S^{\text{KPZ}}(h,T)}{\Delta}\right].
\label{kpdis}
\end{eqnarray}
The height field $h$ and noise field $\tilde p$ are canonically
conjugate variables in the canonical phase space $(\{h({\bf
r}\},\{\tilde p({\bf r}\})$,
\begin{eqnarray}
\{h({\bf r}),\tilde p({\bf r}')\}=\delta^d({\bf r}-{\bf r}').
\label{kpzb}
\end{eqnarray}
We, moreover, have the generators of time translation and space
translation, i.e., energy and momentum,
\begin{eqnarray}
&&H^{\text{KPZ}}=\int d^dx~{\cal H}^{\text{KPZ}}~,~~
\{h,H^{\text{KPZ}}\}=\frac{\partial h}{\partial t},~~ \label{kpzE}
\\
&&{\bm\Pi}^{\text{KPZ}}=\int d^dx~h{\bm\nabla}\tilde
p~,~~\{h,{\bm\Pi}^{\text{KPZ}}\}={\bm\nabla} h. \label{kpzP}
\end{eqnarray}
The Galilean invariance of the KPZ equation (\ref{kpz}) implies
that the noise field $\tilde p$ is invariant; this also follows
from the invariance of the action in Eq.(\ref{kpzact}).

This completes the formal application of the weak noise scheme to
the KPZ equations. The resulting classical field theory must then
be addressed in order to eventually evaluate transition
probabilities and correlations.
\subsection{The Cole-Hopf case}
The Cole-Hopf equation (\ref{che}) is obtained by applying the
nonlinear Cole-Hopf transformation (\ref{ch}) to the KPZ equation
(\ref{kpz}). The Cole-Hopf equation is driven by multiplicative
noise and most work has been based on the mapping to directed
polymers in a random medium \cite{Halpin95}. In the present
context it turns out that a weak noise representation provides a
particular symmetric formulation. Hence, making the assignment,
$w_n(t)\rightarrow w({\bf r},t),p_n(t)\rightarrow p({\bf
r},t),K_{nm}\rightarrow (\lambda/2\nu)^2w^2\delta^d ({\bf r}-{\bf
r}'),F_n\rightarrow -2\left[\nu{\bm\nabla}^2
w-(\lambda/2\nu)wF\right]$, we obtain the Cole-Hopf Hamiltonian
density
\begin{eqnarray}
{\cal H}^{\text{CH}}=p[\nu{\bm\nabla}^2-\nu
k^2]w+\frac{1}{2}k_0^2(wp)^2, \label{chham}
\end{eqnarray}
where we have introduced two characteristic inverse length scales
\begin{eqnarray}
k=\left(\frac{\lambda
F}{2\nu^2}\right)^{1/2}~,~~k_0=\frac{\lambda}{2\nu}.\label{k}
\end{eqnarray}
The field equations are given by
\begin{eqnarray}
&&\frac{\partial w}{\partial
t}=\nu[{\bm\nabla}^2w-k^2w]+k_0^2w^2p, \label{cheq1}
\\
&&\frac{\partial p}{\partial
t}=-\nu[{\bm\nabla}^2p-k^2p]-k_0^2p^2w, \label{cheq2}
\end{eqnarray}
and the reduced action and distribution by
\begin{eqnarray}
&&S^{\text{CH}}(w,T)=\frac{1}{2}k_0^2\int^{w,T}dt d^dx~(w({\bf
r},t)p({\bf r},t))^2,~~~~ \label{chact}
\\
&&P^{\text{CH}}(w,T)\propto
\exp\left[-\frac{S^{\text{CH}}(w,T)}{\Delta}\right].\label{chdis}
\end{eqnarray}
The Cole-Hopf and noise fields $w$ and $p$, spanning the canonical
phase space $(\{w({\bf r})\},\{p({\bf r})\})$, satisfy the Poisson
bracket
\begin{eqnarray}
\{w({\bf r}),p({\bf r}')\}=\delta^d({\bf r}-{\bf r}'), \label{chb}
\end{eqnarray}
For the total energy $H^{\text{CH}}$ and momentum ${\bm
\Pi}^{\text{CH}}$ we have
\begin{eqnarray}
&&H^{\text{CH}}=\int d^dx~{\cal H}^{\text{CH}}~,~~
\{w,H^{\text{CH}}\}=\frac{\partial w}{\partial t}, \label{chE}
\\
&&{\bm \Pi}^{\text{CH}}=\int d^dx~w{\bm\nabla}p~,~~\{w,{\bm
\Pi}^{\text{CH}}\}={\bm\nabla}w. \label{chP}
\end{eqnarray}
Finally, since the action (\ref{chact}) is invariant under the
Galilean transformations (\ref{gal1}) and (\ref{gal5}) the noise
field $p$ must transform according to
\begin{eqnarray}
p\rightarrow p\exp\left[-(\lambda/2\nu){\bf u}^0\cdot{\bf
r}\right].\label{gal6}
\end{eqnarray}
%
\subsection{The Burgers case}
The noisy Burgers equation (\ref{bur}) for the noise-driven slope
of an interface will also be needed in our weak noise analysis. In
this case we choose the assignment: $w_n(t)\rightarrow {\bf
u}({\bf r},t),p_n(t)\rightarrow {\bf p}({\bf
r},t),K_{nm}\rightarrow{\bm\nabla}^2\delta^d ({\bf r}-{\bf
r}'),F_n\rightarrow -2[\nu{\bm\nabla}^2 {\bf u}+\lambda({\bf
u}\cdot{\bm\nabla}){\bf u}]$. Hence, the Hamiltonian density is
\begin{eqnarray}
{\cal H}^{\text{B}}={\bf p}\left[\nu{\bm\nabla}^2{\bf
u}+\lambda({\bf u}\cdot{\bm\nabla}){\bf
u}-\frac{1}{2}{\bm\nabla}({\bm\nabla}\cdot{\bf p})\right],
\label{burham}
\end{eqnarray}
and the ensuing field equations
\begin{eqnarray}
&&\left(\frac{\partial}{\partial t}-\lambda{\bf
u}\cdot{\bm\nabla}\right){\bf u}=\nu{\bm\nabla}^2{\bf
u}-{\bm\nabla}({\bm\nabla}\cdot{\bf u}), \label{bureq1}
\\
&&\left(\frac{\partial}{\partial t}-\lambda{\bf
u}\cdot{\bm\nabla}\right){\bf p}= \nonumber
\\
&&-\nu{\bm\nabla}^2{\bf p}+\lambda({\bf p}({\bm\nabla}\cdot{\bf
u})-({\bf p}\cdot{\bm\nabla}){\bf u}),\label{bureq2}
\end{eqnarray}
where we have used the property that ${\bf u}$ is a longitudinal
vector field. Since the operator $(\partial/\partial t-\lambda{\bf
u}\cdot{\bm\nabla})$ is invariant under the Galilei
transformations (\ref{gal1}) and (\ref{gal4}) the equations of
motion (\ref{bureq1}) and (\ref{bureq2}) are manifestly Galilean
invariant with an invariant noise field ${\bf p}$.

For the reduced action and distribution we have
\begin{eqnarray}
&&S^{\text{B}}({\bf u},T)=\frac{1}{2}\int^{{\bf u},T}dtd^dx~{\bf
p}({\bf r},t)^2, \label{buract2}
\\
&&P^{\text{B}}({\bf u},T)= \exp\left[-\frac{S^{\text{B}}({\bf
u},T)}{\Delta}\right].\label{burdis}
\end{eqnarray}
For the fields ${\bf u}$ and ${\bf p}$, spanning the canonical
phase space $(\{{\bf u}({\bf r})\},\{{\bf p}({\bf r})\})$, we
obtain the Poisson bracket
\begin{eqnarray}
\{u_n({\bf r}),p_m({\bf r}')\}=\delta_{nm}\delta^d({\bf r}-{\bf
r}'). \label{burb}
\end{eqnarray}
Moreover, the total energy $H^{\text{B}}$ and momentum ${\bm
\Pi}^{\text{B}}$ are given by
\begin{eqnarray}
&&H^{\text{B}}=\int d^dx~{\cal H}^{\text{B}}~,~~ \{{\bf
u},H^{\text{B}}\}=\frac{\partial{\bf u}}{\partial t}, \label{burE}
\\
&&{\bm \Pi}^{\text{B}}=\int d^dx~u_n{\bm\nabla}p_n~,~~\{u_n,{\bm
\Pi}^{\text{CH}}\}={\bm\nabla}u_n.~~~~ \label{burP}
\end{eqnarray}
We note that the action (\ref{buract2}) also implies that the
noise field ${\bf p}$ is invariant under the Galilean
transformation.
\subsection{Phase space behavior and canonical transformations}
In the KPZ, Cole-Hopf, and Burgers cases the orbits or
trajectories from an initial configuration $(h^{\text{i}},
w^{\text{i}},{\bf u}^{\text{i}})$ to a final configuration $(h,
w,{\bf u})$ traversed in time $T$, yielding the actions and
transition probabilities, live in the corresponding phase space
spanned by the canonically conjugate fields, i.e., the height
field $h$, the diffusive Cole-Hopf field $w$, the Burgers slope
field ${\bf u}$ and their associated noise fields. The canonical
field theories are conserved and the orbits are confined to
constant energy manifolds, $H^{\text{KPZ}}=\text{const.}$,
$H^{\text{CH}}=\text{const.}$, and $H^{\text{B}}=\text{const}$.

The structure of phase space determines the nature of the
underlying stochastic model. Here the zero-energy manifolds,
$H^{\text{KPZ}} = H^{\text{CH}}= H^{\text{B}}=0$, play a central
role in determining the stationary stochastic state. First we
notice that in all three cases a vanishing  noise field $p=0$,
$\tilde p=0$, and ${\bf p}=0$ is consistent with the field
equations (\ref{kpzeq1}), (\ref{kpzeq2}), (\ref{cheq1}),
(\ref{cheq2}), (\ref{bureq1}), and (\ref{bureq2}), and yield
$H^{\text{KPZ}} = H^{\text{CH}}= H^{\text{B}}=0$. On the transient
zero-noise submanifold we thus obtain the deterministic damped
evolution equations
\begin{eqnarray}
&&\frac{\partial h}{\partial t}=\nu{\bm\nabla}^2h
+\frac{\lambda}{2}{\bm\nabla} h\cdot{\bm\nabla} h-F,
\label{kpzdet}
\\
&&\frac{\partial w}{\partial t} =\nu[{\bm\nabla}^2w-k^2w]~,
\label{chdet}
\\
&&\frac{\partial{\bf u}}{\partial t} =\nu{\bm\nabla}^2{\bf
u}+\lambda({\bf u}\cdot{\bm\nabla}){\bf u}, \label{burdet}
\end{eqnarray}
describing the transient relaxation of the height, Cole-Hopf, and
Burgers fields subject to transient pattern formation. On the
transient zero-energy submanifold $w$ vanishes in the long wave
length limit as $w\propto\exp[-\nu k^2t]$, corresponding to
$h=(2\nu/\lambda)\log w\sim -Ft$, i.e., setting the drift $F=0$ by
the transformation $h\rightarrow h-Ft$, the height field likewise
vanishes as does the slope field ${\bf u}$.

The linear Cole-Hopf equation (\ref{chdet}) is exhausted by
diffusive modes, i.e.,
\begin{eqnarray}
w({\bf r},t)=e^{-\nu k^2 t}\int\frac{d^dp}{(2\pi)^d}~A({\bf
p})e^{-\nu p^2t}e^{i{\bf pr}}, \label{chdetsol}
\end{eqnarray}
where $A({\bf p})$ is chosen according to the imposed initial and
boundary conditions. Together with the relations $h=(1/k_0)\log w$
and ${\bf u} = (1/k_0){\bm\nabla}w/w$ this constitutes a complete
solution of the deterministic transient case.

In 1D the transient pattern formation in the slope field is
composed of propagating right hand domain walls  connected by
ramps with super imposed linear diffusive modes. For the height
profile this corresponds to a pattern of upward pointing cusps
(domain walls) connected by parabolic segments (ramps)
\cite{Fogedby98a}. For further analysis in higher dimensions see
e.g. Ref. \cite{Woyczynski98}.

In the presence of noise, corresponding to the coupling to the
noise fields ($p$, $\tilde p$, or ${\bf p}$), the orbits in phase
space from an initial configuration at time $t=0$ to a final
configuration at time $T$ with the noise field as a slaved
variable, veers away from the transient manifold, pass close to a
saddle point, and asymptotically approach another zero-energy
submanifold intersecting the transient submanifold at the saddle
point. This behavior follows from the general discussion in
Sec.~\ref{weak} and is shown in the case of the overdamped
oscillator discussed in appendix \ref{app}.

The noisy or stationary zero-energy submanifold determines the
stationary stochastic state. In the limit $T\rightarrow\infty$ the
orbit from fixed initial to final configurations migrate to the
zero-energy manifold, moves along the transient submanifold,
passes through the saddle point experiencing a long (infinite)
waiting time, and eventually moves along the stationary
submanifold to the final configuration. The infinite waiting time
at the saddle point thus ensures Markovian behavior, i.e., loss of
memory, in that the stationary distribution only depends on the
final configuration in the limit $T\rightarrow\infty$.

The stationary submanifold is determined by the orthogonality
condition, for example, in the KPZ case,
\begin{eqnarray}
&&H^{\text{KPZ}}= \nonumber \\
&&\int d^dx~\tilde
p\left[\nu{\bm\nabla}^2h+\frac{\lambda}{2}{\bm\nabla}h\cdot{\bm\nabla}
h -F+\frac{1}{2}\tilde p\right]=0,~~~~\label{kpzortho}
\end{eqnarray}
i.e., the manifold $\nu{\bm\nabla}^2h+(\lambda/2){\bm\nabla}
h\cdot{\bm\nabla}h-F+(1/2)\tilde p$ orthogonal to $\tilde p$. In
the case of the overdamped oscillator, discussed in appendix
\ref{app}, for one degree of freedom defined by Eq.
(\ref{oscham}), $H=(1/2)p(p-2\gamma x)$, the zero-energy
submanifolds are $p=0$ and $p=2\gamma x$. The action with $H=0$
then immediately yields $S=\int dt~pdx/dt=\gamma x^2$ and hence
the stationary distribution in Eq. (\ref{oscstat}). On the other
hand, in the field theoretical case characterized by an infinite
numbers of degrees of freedom it is in general a difficult task to
determine the manifold $p=F(h)$ satisfying the orthogonality
condition (\ref{kpzortho}).

As discussed in Sec.~\ref{kpzeq} the situation is special in 1D.
Here the stationary Fokker-Planck equation in the KPZ case admits
an explicit solution given by Eq. (\ref{stakpz}) \cite{Huse85}.
Within the weak noise scheme the existence of this
fluctuation-dissipation theorem is tantamount to the explicit
determination of the stationary submanifold. This is most easily
done in the Burgers case with Hamiltonian density ${\cal
H}^{\text{B}}=p(\nu\nabla^2 u+\lambda u\nabla u-(1/2)\nabla^2 p)$.
Setting $\nabla^2p=2\nu\nabla^2 u$ we have ${\cal
H}^{\text{B}}=(2\nu\lambda/3)\nabla(u^3)$, i.e., a total
derivative yielding $H^{\text{B}}=0$ with vanishing slope boundary
conditions. For the action we obtain $S=\int dt dx2\nu u\partial
u/\partial t=\nu\int dx u^2=\nu\int dx(\nabla h)^2$ yielding the
stationary distribution in Eq. (\ref{stakpz}), see also Refs.
\cite{Fogedby99a,Fogedby03b}.

The stochastic KPZ, Cole-Hopf, and Burgers equations (\ref{kpz}),
(\ref{che}), and (\ref{bur}) all constitute equivalent
descriptions of a growing interface. Within the weak noise
approach the canonical structure implies that the three equivalent
descriptions are related by canonical transformations. By
inspection we find that the Cole-Hopf and KPZ formulations are
connected by the canonical transformations
\cite{Goldstein80,Landau59b}
\begin{eqnarray}
w=\exp[k_0h]~,~~ p=k_0^{-1}\tilde p\exp[-k_0 h], \label{chkpz}
\end{eqnarray}
together with the inverse transformations
\begin{eqnarray}
h=k_0^{-1}\log w~,~~ \tilde p=k_0 pw; \label{kpzch}
\end{eqnarray}
the generating function is $G_1(p,h)=p\exp[k_0h]$, i.e.,
$dG_1=wdp+\tilde pdh$, implying $w=\partial G_1/\partial p$ and
$\tilde p=\partial G_1/\partial h$. Likewise, the KPZ and Burgers
formulations are related by means of the transformations
\begin{eqnarray}
{\bf u}={\bm\nabla}h~,~~\tilde p=-{\bm\nabla}\cdot{\bf p},
\label{burkpz}
\end{eqnarray}
with generating function $G_2(h,{\bf p})={\bf
p}\cdot{\bm\nabla}h$, i.e., $dG_2=\tilde p dh+{\bm\nabla}h\cdot
d{\bf p}$, implying $\tilde p=\partial G_2/\partial h$, and ${\bf
u}=\partial G_2/\partial{\bf p}$.

From the field equations in the Burgers case, Eqs. (\ref{bureq1})
and (\ref{bureq2}), it follows that only the longitudinal
component of ${\bf p}$ couples to the slope field ${\bf u}$. Since
Eq. (\ref{bureq2}) is linear in ${\bf p}$ we can without loss of
generality just keep the longitudinal component of ${\bf p}$,
i.e., ${\bf p}={\bm\nabla}\phi$, where $\phi$ is a scalar
potential. We thus obtain the canonical transformation ${\bf
u}={\bm\nabla}h$ and $\tilde p=-\nabla^2\phi$ and the field
equation (\ref{bureq1}) takes the form $(\partial/\partial
t-\lambda{\bf u}\cdot{\bm\nabla}){\bf u}=\nu\nabla^2{\bf
u}+\nabla^2\phi$.
%
\section{\label{field} Field equations}
The canonically conjugate field equations of motion in the KPZ,
Cole-Hopf, or Burgers formulations form the central starting point
for an analysis of the pattern formation and scaling properties of
the KPZ equation. As discussed above the three formulations are
related by canonical transformations and represent the same
stochastic problem in the weak noise limit.
%
\subsection{General properties}
The field equations, determining orbits in the corresponding
multi-dimensional phase space, have the form of coupled nonlinear
hydrodynamical equations. Common to all three formulations is that
the field equations for the noise fields $p$,$\tilde p$, or ${\bf
p}$ have negative diffusion coefficients. This corresponds to a
Fourier mode of the noise field growing exponentially in time,
rendering a numerical integration forward in time unfeasible
\cite{Fogedby02a}. The growth of the noise field is consistent
with the property that the noise drives the system into a
stationary state at long times corresponding to orbits in phase
space leaving the transient submanifold, see e.g. the discussion
of the overdamped oscillator in appendix \ref{app}.

Leaving aside the possibility that the coupled field equations in
the KPZ, Cole-Hopf, or Burgers formulations are exactly integrable
in the sense that a Lax pair and an inverse scattering
transformation can be identified, see e.g. Ref.
\cite{Fogedby80,Scott99,Jackson90}, we note that the equations of
motion all are invariant subject to a Galilei transformation
combined with a rescaling or shift of the fields. This property
suggest the possibility of constructing localized propagating
solitons or elementary excitations by first finding static
localized solutions which subsequently are boosted by a Galilean
transformation.
\subsection{The one dimensional Burgers case}
\begin{figure}
\includegraphics[width=1.0\hsize]
{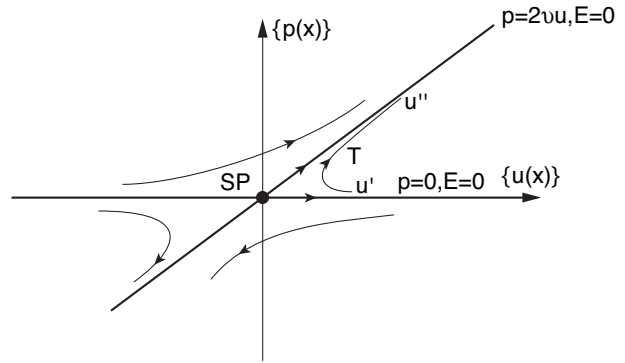} \caption{We show the phase space structure in the 1D
Burgers case. The transient $E=0$, $p=0$ submanifold and the
stationary $E=0$, $p=2\nu u$ submanifold intersect at the saddle
point SP. We depict a finite time orbit from $u'$ to $u''$ in time
$T$.} \label{fig4}
\end{figure}

In the 1D case summarized in detail in Ref. \cite{Fogedby03b} the
scalar slope field $u=dh/dx$ turns out to be the convenient
variable. Since this case has served as a theoretical laboratory
we review it here. In the 1D Burgers case the Hamiltonian takes
the form
\begin{eqnarray}
H^{\text{B}}=\int dx~p[\nu\nabla^2u+\lambda u\nabla
u-(1/2)\nabla^2p], \label{burham1}
\end{eqnarray}
yielding the equations of motion
\begin{eqnarray}
&&\left(\frac{\partial}{\partial t}-\lambda
u\nabla\right)u=\nu\nabla^2u-\nabla^2p, \label{bur1eq1}
\\
&&\left(\frac{\partial}{\partial t}-\lambda
u\nabla\right)p=-\nu\nabla^2p; \label{bur1eq2}
\end{eqnarray}
note that the last term in Eq. (\ref{bureq2}) vanishes in 1D. In
Fig.~\ref{fig4} we have depicted the Burgers phase space
structure.
%
\subsubsection{Domain wall solutions and pattern formation in 1D}
\begin{figure}
\includegraphics[width=1.0\hsize]
{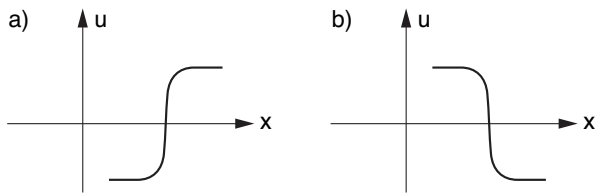} \caption{We show the fundamental static right hand and
left hand domain wall or soliton solutions for the 1D Burgers
equation. The right hand domain wall is a solution in the
deterministic case; the left hand domain wall is induced by the
noise.}\label{fig5}
\end{figure}

\begin{figure}
\includegraphics[width=.8\hsize]
{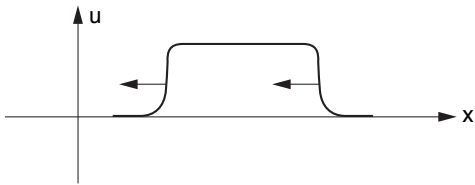} \caption{We show the propagating two-domain wall quasi
particle or excitation for the 1D Burgers equation.}\label{fig6}
\end{figure}

In addition to superimposed extended phase-shifted diffusive modes
with dispersion $\omega=\pm\nu p^2$, the equations (\ref{bur1eq1})
and (\ref{bur1eq2}) support two localized distinctive soliton or
domain wall modes, in the static case of the kink-like form,
\begin{eqnarray}
u(x)=\pm\frac{k}{k_0}\tanh kx, \label{statbur1}
\end{eqnarray}
with inverse scales $k$ and $k_0$ given by Eq. (\ref{k}). Boosting
a pair of well-separated matched right and left hand domain walls
by means of the Galilean transformation (\ref{gal1}) and
(\ref{gal4}) to the velocity $v=2\nu k$, corresponding to the
slope shift $u^0=k/k_0$, we obtain a propagating elementary
excitation or quasi particle with vanishing $u$ at infinity. The
moving domain wall pair is equivalent to a moving step in the
height profile, corresponding to adding a layer to the growing
interface at each passage of the quasi particle. In
Fig.~\ref{fig5} we have shown the fundamental static right hand
and left hand domain wall solutions. In Fig.~\ref{fig6} we have
depicted the two-domain wall quasi particle.

A general growth morphology is obtained by matching a dilute gas
of propagating domain walls in terms of the slope field; for the
height field this morphology corresponds to a growing interface.
Superimposed on the domain wall pattern are extended diffusive
modes. In the limit of vanishing nonlinearity for $\lambda=0$ the
domain wall gas vanishes, the growth ceases, and the diffusive
excitations exhaust the mode spectrum.

\begin{figure}
\includegraphics[width=1.0\hsize]
{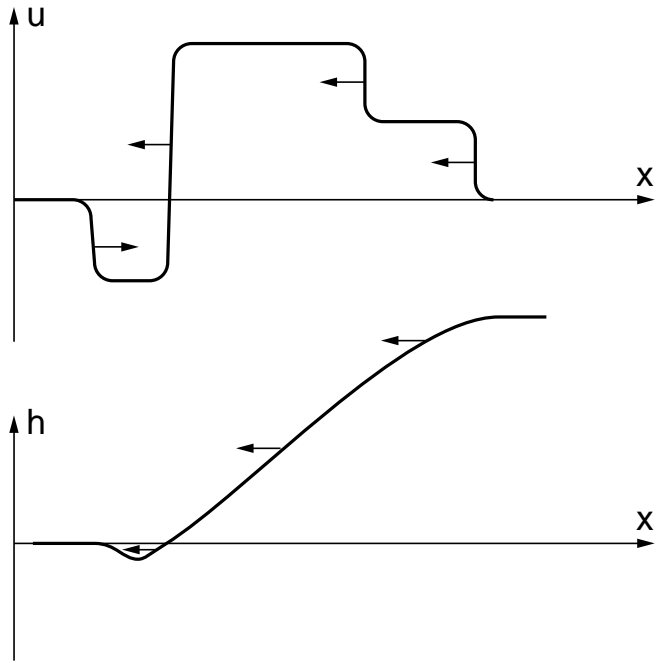} \caption{We show a multi-domain wall representation of a
growing interface in the slope field and the associated height
field.}\label{fig7}
\end{figure}

In terms of the height field $h$ the static domain wall solutions
(\ref{statbur1}) corresponds to the profiles
\begin{eqnarray}
h(x)=\pm\frac{1}{k_0}\ln|\cosh kx|, \label{statkpz1}
\end{eqnarray}
i.e., concave and convex cusps. The corresponding Cole-Hopf field
$w$ is given by
\begin{eqnarray}
w(x)=\cosh^{\pm 1} kx, \label{statch1}
\end{eqnarray}
corresponding to a localized bound state of width $1/k$ falling
off like $w(x)\propto\exp(-k|x|)$ and a concave profile increasing
like $w(x)\propto\exp(k|x|)$. In Fig.~\ref{fig7} we have depicted
a multi-domain wall representation of a growing interface and the
associated height profile.

%
\subsubsection{Scaling properties in 1D}
The scaling properties follow from the dynamics of the domain
walls. The right hand domain wall corresponds to vanishing noise
field $p=0$ and is the well-known viscosity-smoothed shock wave of
the noiseless deterministic Burgers equation
\cite{Burgers29,Fogedby98a}. The left hand domain wall, on the
other hand, is associated with the noise field $p=2\nu u$ and
carries action $S=(1/6)\nu\lambda u^3T$, energy
$E=-(1/6)\nu\lambda u^3$, and momentum $\Pi=0$, where $u=2k/k_0$
is the domain wall amplitude. For the quasi particle composed of a
right hand and left hand domain wall moving with velocity $v=2\nu
k$ we obtain $S=(1/6)\nu\lambda u^3 T$, $E=-(2/3)\nu\lambda u^3$,
and $\Pi=\nu u^2$. Eliminating the amplitude the quasi particle is
characterized by the gapless dispersion law
\begin{eqnarray}
E=-\frac{4}{3}\frac{\lambda}{\nu^{1/2}}|\Pi|^{3/2},
\label{soldisp1}
\end{eqnarray}
with power law exponent $3/2$. Note that the diffusive modes have
the dispersion law $\omega\propto\nu k^2$, corresponding to the
power law exponent $2$.

The scaling exponents $z=3/2$ and $\zeta=1/2$ follow from i) a
spectral representation of the slope correlations and ii) the
structure of the zero energy manifolds. i) Drawing on the analogy
with a quantum system with Planck constant $\Delta$ we invoke
heuristically a spectral representation for the slope correlations
\begin{eqnarray}
\langle uu\rangle(x,t)=\int d\Pi F(\Pi)e^{-\frac{E}{\Delta}t+i\Pi
x}, \label{spec1}
\end{eqnarray}
where $F(\Pi)$ is an appropriate form factor. Within the single
quasi particle sector, inserting the dispersion law
(\ref{soldisp1}), it readily follows that $t$ scales with
$x^{3/2}$ and we infer the dynamic scaling exponent $z=3/2$. ii)
From the Hamiltonian (\ref{burham1}) we infer the zero-energy
manifolds $p=0$ and $p=2\nu u$, consistent with the equations of
motion (\ref{bur1eq1}) and (\ref{bur1eq2}). At long times the
orbits in the $(u,p)$ phase space approach the zero-energy
manifold $p=2\nu u$ and we obtain for the action $S\sim \int dx dt
p\partial u/\partial t=2\nu\int dx dt u\partial u/\partial t =
\nu\int dx u^2$,i.e., Eq. (\ref{stakpz}), and the independent
stationary fluctuations of $u$ are given by a Gaussian
distribution. This in turn implies that the height field $h=\int^x
dx' u$ performs random walk yielding according to the scaling form
(\ref{corkpz}) the roughness exponent $\zeta = 1/2$. Note that the
scaling law $z+\zeta =2$ is automatically obeyed since the weak
noise formulation is consistently Galilean invariant.

The notion of universality classes is here associated with the
dominant gapless quasi particle dispersion law. The scaling
properties follow from the low frequency (long time) - small
wavenumber (large distances) limit. For $\lambda =0$ there is no
growth, the mode spectrum is exhausted by extended diffusive modes
with gapless dispersion $\omega=\nu k^2$, yielding according to
the spectral form the dynamic exponent $z=2$. The roughness
exponent $\zeta=1/2$ and the scaling law is not operational. This
constitutes the Edwards-Wilkinson universality class. For
$\lambda\neq 0$ localized domain wall growth modes nucleate out of
the diffusive mode continuum with dispersion
$E\propto(\lambda/\nu^{1/2})\Pi^{3/2}$; this is the KPZ
universality class.

\begin{figure}
\includegraphics[width=1.0\hsize]
{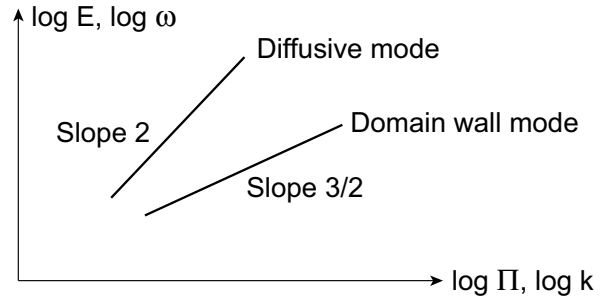}\caption{We show in a log-log plot the domain wall
dispersion law $E\propto\Pi^{3/2}$ and the diffusive mode
dispersion law $\omega\propto k^2$.} \label{fig8}
\end{figure}

Summarizing, in 1D the growing interface problem can be analyzed
in some detail and a consistent interpretation in the WKB sense
can be advanced within the weak noise formulation. The approach
yields: i) a many body description of a growing interface in terms
of a 1D matched network of moving domain walls with superimposed
diffusive modes, ii) scaling properties and scaling exponents
follow from the dispersion law of the dominant gapless domain wall
excitations and the structure of the zero-energy manifold in phase
space, iii) universality classes are associated with the class of
gapless excitations governing the dynamics of the interface. In
Fig.~\ref{fig8} we have in a log-log plot depicted the domain wall
and diffusive mode dispersion laws.
%
\subsection{The Cole-Hopf case}
In higher dimension we must address the field equations of motion
in either the KPZ formulation, Eqs. (\ref{kpzeq1}) and
(\ref{kpzeq2}), the Cole-Hopf formulation, Eqs. (\ref{cheq1}) and
(\ref{cheq2}), or the Burgers formulation, Eqs. (\ref{bureq1}) and
(\ref{bureq2}). Based on the working hypothesis that the growth
morphology is associated with a network of growth modes and
drawing from the insight gained in 1D, the program is again to
search for localized solutions of the equations of motion. Whereas
both the KPZ and Burgers formulations do not easily yield to
analysis, the symmetric Cole-Hopf formulation turns out to be the
convenient starting point
\subsubsection{Static localized modes}
In the static limit  the Cole-Hopf equations of motion
(\ref{cheq1}) and (\ref{cheq2}) assume the symmetrical form
\begin{eqnarray}
&&\nu{\bm\nabla}^2 w=\nu k^2w-k_0^2 w^2p, \label{statcha}
\\
&&\nu{\bm\nabla}^2 p=\nu k^2p-k_0^2 p^2w. \label{statchb}
\end{eqnarray}
They are the Euler equations determining the configurations
associated with the extrema of the Cole-Hopf Hamiltonian
(\ref{chE}), i.e., $\delta H^{\text{CH}}/\delta w=0$ and $\delta
H^{\text{CH}}/\delta p= 0$. By inspection we note that the Euler
equations are compatible for $p=0$ (and $w=0$) and for $p\propto
w$. For $p=0$ Eq. (\ref{statchb}) is satisfied identically, for
$p=\nu w$ Eqs. (\ref{statcha}) and (\ref{statchb}) are identical;
the prefactor $\nu$ is dictated by dimensional arguments.

On the noiseless manifold $p=0$ with $H^{\text{CH}}=0$ we obtain
the linear (Helmholtz-type) equation
\begin{eqnarray}
{\bm\nabla}^2w=k^2 w. \label{chlin}
\end{eqnarray}
An elementary solution of Eq. (\ref{chlin}) is $w\propto\exp(k{\bf
\hat e}\cdot{\bf r})$, where ${\bf \hat e}$ is a unit vector,
${\bf\hat e}^2=1$, pointing in an arbitrary direction. This mode
corresponds to the height field $h=(k/k_0){\bf\hat e}\cdot{\bf
r}$, i.e., an inclined plane, and the constant slope field ${\bf
u}=(k/k_0){\bf\hat e}$ of magnitude $k/k_0$ pointing in direction
${\bf \hat e}$. A general solution of Eq. (\ref{chlin}) is
constructed according to $w({\bf r})=\int d^d{\bf\hat e}~
A({\bf\hat e})\exp(k{\bf\hat e}\cdot{\bf r})$, where since $w>0$
we must choose $A({\bf\hat e}) >0$. For the height field and slope
field we obtain correspondingly $h({\bf r})=(1/k_0)\log w({\bf
r})$ and ${\bf u}({\bf r})=(k/k_0){\bm\nabla}w({\bf r})/w({\bf
r})$, respectively. By choosing the weight function $A({\bf\hat
e})$ appropriately we can prescribe the directional dependence of
the exponential growth of $w$ and the corresponding form of the
height and slope fields.

In the later analysis of the network solution it turns out that
rotationally invariant solution of Eq. (\ref{chlin}), i.e., an
s-wave state, will be important in implementing the long distance
boundary conditions. In polar coordinates, ignoring angular
dependence, Eq. (\ref{chlin}) takes the form
\begin{eqnarray}
\frac{d^2w}{dr^2}+\frac{d-1}{r}\frac{dw}{dr}=k^2w.
\label{chradlin}
\end{eqnarray}
At long distances, ignoring the first order term, we have
$w\propto\exp(\pm kr)$. Incorporating the first order term by
setting $w\propto r^\alpha\exp(\pm kr)$ and choosing the growing
solution we obtain
\begin{eqnarray}
w_+^0(r)\sim r^{1/2-d/2}\exp(kr),~~r\rightarrow\infty.
\label{linasymp}
\end{eqnarray}
At small distances and in order to obtain a finite $w$ at $r=0$ we
must choose $dw/dr=\text{cst.}\times r$, implying
\begin{eqnarray}
w_+^0(r)-w_+^0(0)\propto r^2,~~r\rightarrow 0. \label{linorig}
\end{eqnarray}
The exact solution of Eq. (\ref{chradlin}), finite at the origin,
is given by $w_+^0(r)\propto r^{1-d/2}I_{d/2-1}(kr)$, where
$I_\nu(z)$ is the Bessel function of the second kind
\cite{Lebedev72}.

Correspondingly, the asymptotic height and slope fields have the
form
\begin{eqnarray}
h_+^0(r)\sim\frac{k}{k_0}r~,~~{\bf u}_+^0(r)\sim
\frac{k}{k_0}\frac{{\bf r}}{r},~~r\rightarrow\infty,
\label{linasymphu}
\end{eqnarray}
and at small distances
\begin{eqnarray}
h_+^0(r)\propto r^2~,~~{\bf u}_+^0(r)\propto {\bf
r},~~r\rightarrow 0. \label{linorighu}
\end{eqnarray}
At large distances the height field forms a d-dimensional cone
which at short distances becomes a d-dimensional paraboloid. The
slope field has the form of an outward-pointing vector field of
constant magnitude $k/k_0$; for small $r$ the slope field vanishes
like ${\bf r}$. In 1D $I_{-1/2}(kx)=(2/\pi kx)^{1/2}\cosh kx$ and
we obtain the Cole-Hopf, height, and slope fields in Eqs.
(\ref{statch1}), (\ref{statkpz1}), and (\ref{statbur1}), i.e., the
fields pertaining to the right hand domain wall solutions for the
noiseless Burgers equation.

On the noisy manifold $p=\nu w$ we obtain the stationary nonlinear
Schr\"odinger equation (NLSE)
\begin{eqnarray}
\nabla^2 w=k^2 w-k_0^2 w^3, \label{chnlse}
\end{eqnarray}
which can be recognized as the stationary Gross-Pitaevski equation
for a real Bose condensate with energy $k^2$ and coupling strength
$k_0^2$ \cite{Pethick02}. In radial coordinates we have
\begin{eqnarray}
\frac{d^2w}{dr^2}+\frac{d-1}{r}\frac{dw}{dr}=k^2w-k_0^2w^3,
\label{chradnlse}
\end{eqnarray}
which supports a nodeless bound state $w_-^0(r)$ falling off at
large $r$ and finite at the origin \cite{Chiao64}. For small $w$
we recover the linear equation (\ref{chradlin}) with decaying
solution
\begin{eqnarray}
w_-^0(r)\sim r^{1/2-d/2}\exp(-kr),~~r\rightarrow\infty.
\label{nlseasymp}
\end{eqnarray}
For small $r$, requiring finiteness of the first order term, we
infer
\begin{eqnarray}
w_-^0(r)-w_-^0(0)\propto - r^2,~~r\rightarrow 0. \label{nlseorig}
\end{eqnarray}
Correspondingly, the asymptotic height and slope fields have the
form
\begin{eqnarray}
h_-^0(r)\sim-\frac{k}{k_0}r~,~~{\bf u}_-^0(r)\sim
-\frac{k}{k_0}\frac{{\bf r}}{r},~~r\rightarrow\infty,
\label{nlseasymphu}
\end{eqnarray}
and at small distances
\begin{eqnarray}
h_-^0(r)\propto -r^2~,~~{\bf u}_-^0(r)\propto- {\bf
r},~~r\rightarrow 0. \label{nlseorighu}
\end{eqnarray}
At large distances the height field forms an inverted
d-dimensional cone, at short distances an inverted d-dimensional
paraboloid. The slope field forms an inward-pointing vector field
of constant magnitude $k/k_0$; for small $r$ the slope field
vanishes like $-{\bf r}$. In 1D the radial equations
(\ref{chradnlse}) takes the form $d^2w/dr^2=k^2w-k_0^2w^3$ and
admits the solution $w_-^0(x)=(k/k_0)\sqrt 2\cosh^{-1}kx$ in
accordance with Eq. (\ref{statch1}), yielding the height and slope
fields in Eqs. (\ref{statkpz1}) and (\ref{statbur1}), i.e., the
fields for the left-hand domain wall constituting the
noise-induced growth modes for the 1D noisy Burgers equation.

\begin{figure}
\includegraphics[width=1.0\hsize]
{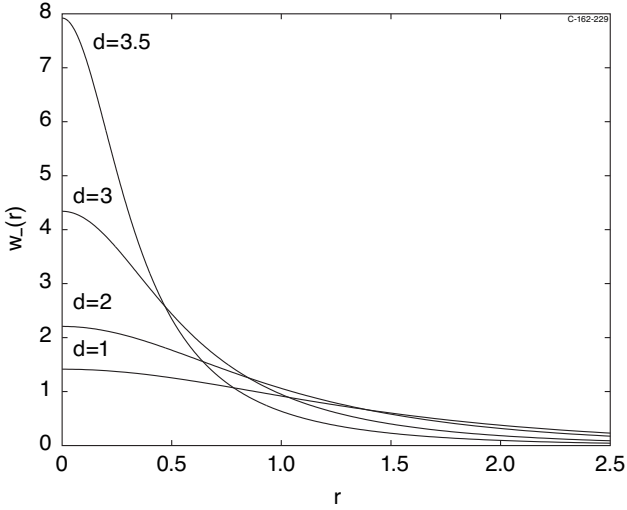}\caption{The radially symmetric bound states of the NLSE
are depicted for $k=k_0=1$ in $d=1, d=2, d=3, d=3.5$. In the limit
$d\rightarrow 4$ the bound state solution disappears.}
\label{fig9}
\end{figure}

\begin{figure}
\includegraphics[width=1.0\hsize]
{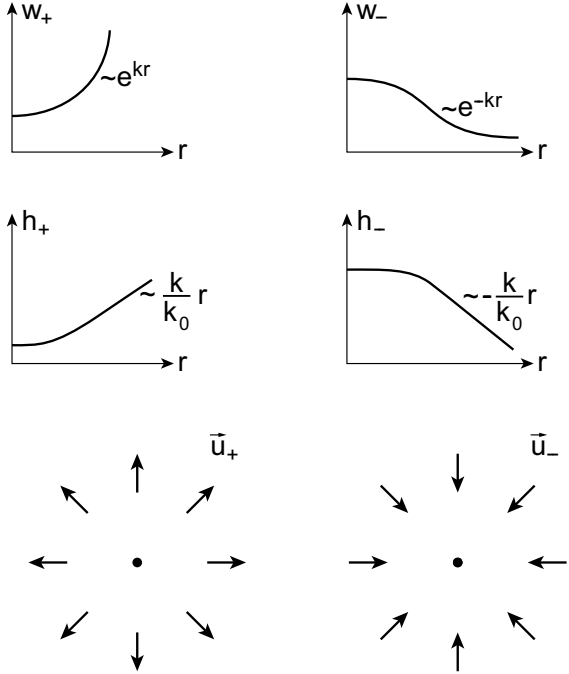}\caption{We show the two static radial modes for the
Cole-Hopf, KPZ, and Burgers fields. The KPZ fields correspond to
cones at large distance and paraboloids at short distances; the
Burgers fields are outward-pointing and inward-pointing vector
fields vanishing linearly at the origin.} \label{fig10}
\end{figure}

By a simple scaling argument, $w\rightarrow \mu w$,
$k_0\rightarrow k_0/\mu$, the coupling strength $k_0$ can be
scaled to $1$. Consequently, the length scale is set by $k^{-1}$.
Generally,
\begin{eqnarray}
w_-^0(r)=a_d\frac{k}{k_0}f(kr), \label{nlsedimless}
\end{eqnarray}
where $f(kr)\sim (kr)^{(1-d)/2}\exp(-kr)$ for $kr\gg 1$.
Normalizing $f(kr)=1$ for $r=0$, i.e., $f(0)=1$, the dimensionless
coefficient $a_d$ is a function of the spatial dimension $d$. In
$d=1$ we have from above $a_1=\sqrt 2$, in higher dimension $a_d$
is determined numerically. We find $a_2=2.21$ and $a_3=4.34$. In
$d\ge 4$ the bound state is absent. This interesting feature will
be discussed later in the context of the upper critical dimension.
In Fig.~\ref{fig9} we have depicted the radially symmetric bound
states of the NLSE for dimensions $d=1$, $d=2$, $d=3$, and
$d=3.5$.
The static localized spherical modes $w_+^0(r)$ and $w_-^0(r)$
constitute the building block in establishing the growth
morphology of the KPZ equation. In 1D they become for the slope
field the right hand and left hand domain wall solutions. In
Fig.~\ref{fig10} we have depicted the two kinds of radial modes
for the $w$-field, $h$-field, and ${\bf u}$-field.

\begin{figure}
\includegraphics[width=1.0\hsize]
{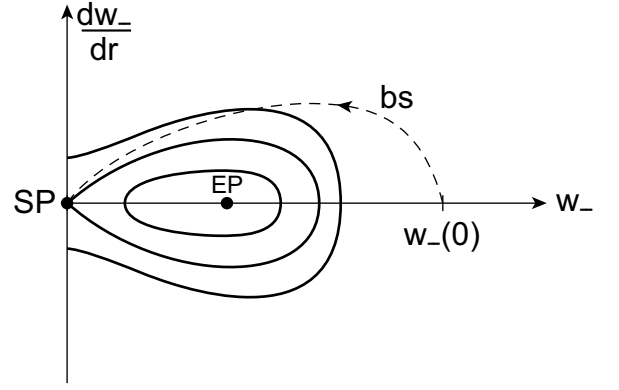}\caption{Phase space plot of $dw_-/dr$ vs. $w_-$. We
depict the constant energy surfaces
$E=(1/2)(dw/dr)^2+V(w)=\text{cst.}$ The dashed line corresponds to
the radial bound state solution of the NLSE.} \label{fig11}
\end{figure}

Following Finkelstein \cite{Finkelstein51} we finally give a
simple argument based on dynamical system theory for the existence
a radially symmetric bound states of the NLSE (\ref{chradnlse}).
Treating $r$ as a time variable Eq. (\ref{chradnlse}) describes
the motion of a particle of unit mass with $w$ as position in the
double well potential $V(w)=-(k^2/2)w^2+(k_0^2/4)w^4$, subject to
the time-dependent damping $(d-1)(dw/dr)/r$. The phase space is
spanned by $w$ and $dw/dr$ and characterized by a saddle point at
$(w,dw/dr)=(0,0)$ and an elliptic point at $(w,dw/dr)=(k/k_0,0)$.
The invariant homoclinic orbit (the separatrix) intersect the $w$
axis at $w=0$ and $w=\sqrt 2 k/k_0$. In Fig.~\ref{fig11} we have
in a phase space plot depicted the constant energy surfaces
$E=(1/2)(dw/dr)^2+V(w)$ in the absence of damping.
In $d=1$ the damping is absent and the bound state solution
$w_-(x)\propto\cosh^{-1}kx$ corresponds to the motion along the
invariant curve, the separatrix, from $w=\sqrt 2 k/k_0$ to the
saddle point. For higher $d$, in the presence of damping, it
follows from the equation of motion that
$dE/dr=-(d-1)(dw/dr)^2/r<0$, i.e., the energy decreases
monotonically in time. The orbit from an initial $w_-(0)$ with
$dw_-/dr|_{r=0}=0$ will intersect the energy surfaces and in most
cases spiral into the elliptic fixed point at $(k/k_0,0)$. Only a
particular orbit, corresponding to the bound state, will reach the
saddle point at $(0,0)$ for $r\rightarrow\infty$. It follows from
the numerical analysis that for increasing dimension, i.e.,
increasing damping, the initial value $w_-(0)$ defining the bound
state solution migrates to larger values and out to infinity for
$d\ge 4$. We have unsuccessfully attempted to determine the
critical dimension by a dynamical system theory argument; however,
in Sec.~\ref{upper} we determine the upper critical dimension
algebraically by means of an application of Derrick's theorem.
%
\subsubsection{Dynamical growth modes}
\begin{figure}
\includegraphics[width=1.0\hsize]
{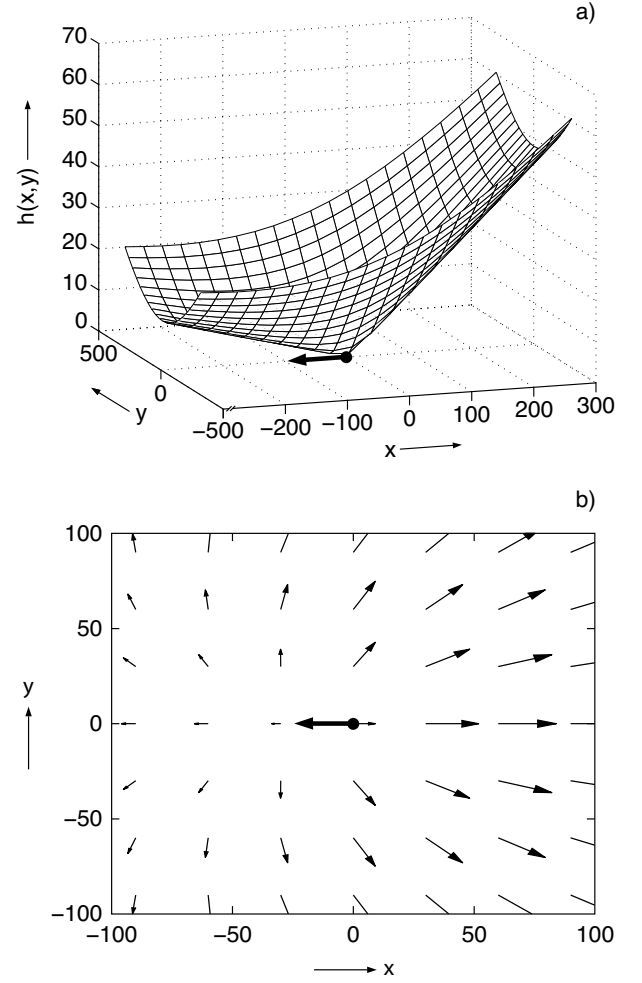} \caption{We depict the propagating height field and slope
field modes in 2D. The bold arrows indicate the propagation
velocity (arbitrary units).} \label{fig12}
\end{figure}

The fundamental building blocks in establishing the growth
morphology of the KPZ equation are the static localized modes
$w_\pm^0(r)$, yielding the height fields $h_\pm^0(r)=(1/k_0)\log
w_\pm^0(r)$ and slope fields ${\bf u}_\pm^0=(1/k_0){\bm\nabla}
w_\pm(r)/w_\pm^0(r)$. The growing mode $w_+^0$ is associated with
the noiseless manifold $p=0$ and carries according to Eqs.
(\ref{chham}), (\ref{chact}), and (\ref{chP}) no dynamics, i.e.,
$S^{\text{CH}}=H^{\text{CH}}=\Pi^{\text{CH}}=0$. The decaying
bound state mode $w_-^0$, on the other hand, is endowed with
dynamical attributes. The mode lives on the noisy manifold $p=\nu
w$ and is associated with the noise field $p_-^0=\nu w_-^0$.
According to Eqs. (\ref{chact}), (\ref{chE}), and (\ref{chP}) it
carries action, energy, and momentum,
\begin{eqnarray}
&&S^{\text{CH}}=\frac{1}{2}(k_0\nu)^2\int d^dxdt~w_-^0(r)^4,
\label{chS1}
\\
&&H^{\text{CH}}=-\frac{1}{2}(k_0\nu)^2 \int d^dx~w_-^0(r)^4,
\label{chE1}
\\
&&{\bm\Pi}^{\text{CH}}=0 \label{chP1}.
\end{eqnarray}
In order to generate dynamical modes we boost the static modes
$w_\pm^0$ by means of the Galilei transformations (\ref{gal1}),
(\ref{gal2}), (\ref{gal4}), and (\ref{gal5}). For the propagating
localized Cole-Hopf, KPZ, and Burgers modes we then obtain
\begin{eqnarray}
&&w_\pm({\bf r},t)=w_\pm^0(|{\bf r}+\lambda{\bf
u}^0t|)\exp(k_0{\bf u}^0\cdot{\bf r}), \label{chdyn}
\\
&&h_\pm({\bf r},t)=h_\pm^0(|{\bf r}+\lambda{\bf u}^0t|)+{\bf
u}^0\cdot{\bf r}, \label{hdyn}
\\
&&{\bf u}_\pm({\bf r},t)={\bf u}_\pm^0(|{\bf r}+\lambda{\bf
u}^0t|)+{\bf u}^0. \label{udyn}
\end{eqnarray}
The height field is a tilted d-dimensional cone moving with
velocity $-\lambda{\bf u}^0$; the slope field  a d-dimensional
hedgehog structure with an imposed constant drift ${\bf u}^0$
moving with velocity $-\lambda{\bf u}^0$. In Fig.~\ref{fig12} we
have shown the height and slope fields in 2D.
Note that in 1D we have $w^0_\pm(x)\propto\cosh^\pm kx$, yielding
$h^0_\pm(x)=(1/k_0)\log|\cosh^\pm kx| + \text{cst.}$ and ${\bf
u}^0_\pm(x)=\pm(k/k_0)\tanh kx$, and we obtain the propagating
modes $u_\pm (x,t)=\pm(k/k_0)\tanh k(x+\lambda u^0 t)+u^0$
discussed in Refs. \cite{Fogedby98b,Fogedby03b}

The dynamics of the propagating modes is easily inferred. The
noiseless mode $(w_+,h_+,{\bf u}_+)$ on the $p=0$ manifold has
vanishing action, energy, and momentum. The noisy mode
$(w_-,h_-,{\bf u}_-)$ on the $p_-=\nu w_-$ manifold carries finite
action, energy, and momentum. Since the action is invariant under
the the Galilean boost the action is given by Eq. (\ref{chS1}).
For the energy and momentum we obtain by insertion in Eqs.
(\ref{chham}) and (\ref{chP}), noting that the noise field is
transformed according to Eq. (\ref{gal6}),
\begin{eqnarray}
H^{\text{CH}}= &&-\frac{1}{2}(k_0\nu)^2\int d^dx~w_-^0(r)^4
\nonumber
\\
&&+(k_0\nu)^2{\bf u}_0^2\int d^dx~w_-^0(r)^2,~~~~~~~~~\label{chE2}
\\
{\bm\Pi}^{\text{CH}}=&&-k_0\nu {\bf u}_0\int d^dx~w_-^0(r)^2.
\label{chP2}
\end{eqnarray}
Since the velocity of the $w_-$ mode is ${\bf v}=-\lambda{\bf
u}^0$ the expressions (\ref{chE2}) and (\ref{chP2}) admit a
particle interpretation. Defining the mass
\begin{eqnarray}
m_-=\frac{1}{2}\int d^dx w_-^0(r)^2, \label{chm}
\end{eqnarray}
we obtain
\begin{eqnarray}
&&{\bm\Pi}^{\text{CH}}=m_-{\bf v}, \label{chP3}
\\
&& H^{\text{CH}}=E_0^{\text{CH}}+(1/2) m_-v^2, \label{chE3}
\end{eqnarray}
where
\begin{eqnarray}
E_0^{\text{CH}}= -(1/2) (k_0\nu)^2\int d^dx~w_-^0(r)^4
\label{chrestE}
\end{eqnarray}
is the rest energy.
\section{\label{pattern} Pattern Formation}
The existence of the localized propagating growth modes in the
Cole-Hopf field and the corresponding configurations in terms of
the height and slope fields makes it natural to describe a growing
interface in terms of a gas of growth modes. The implementation of
this scheme constitutes a generalization of the weak noise
approach in the 1D Burgers case to higher dimensions.
\subsection{General properties}
From the general discussion in Sec.~\ref{weak} and
Sec.~\ref{kpzweak} it follows that a solution of the Cole-Hopf
field equations (\ref{cheq1}) and (\ref{cheq2}) from an initial
configuration $w_{\text{i}}({\bf r})$ at time $t=0$ to a final
configuration $w({\bf r})$ at time $T$, with the noise field
$p({\bf r},t)$ as slaved variable, corresponds to a specific
kinetic pathway, yielding according to Eqs. (\ref{chact}) and
(\ref{chdis}) the corresponding Arrhenius factor. The first
central issue is thus to determine a global solution of the field
equations. In the spirit of instanton calculations in field theory
\cite{Coleman77,Das93,Rajaraman87} and following the scheme
implemented in the case of the noisy 1D Ginzburg-Landau equation
\cite{Fogedby05a} and the noisy 1D Burgers equation
\cite{Fogedby04b} we attempt to build a global solution on the
basis of the propagating localized growth modes. In order to
minimize overlap contributions we, moreover, consider the case of
a dilute gas of growth modes. In order to characterize a kinetic
pathway or growing interface we must, moreover, choose appropriate
boundary conditions. Here it is natural to assume a flat interface
at infinity, that is a vanishing slope field. Note that this
boundary condition does not preclude an offset in the height field
and thus allows for a propagation of facets or textures accounting
for the nonequilibrium growth.
\subsection{Dynamical pair mode}
In order to illustrate how to construct solutions we here consider
a pair mode built from two growth modes. At time $t=0$ we combine
a static noiseless and a static noisy Cole-Hopf mode centered at
${\bf r}_1$ and ${\bf r}_2$, respectively, and obtain the three
equivalent pair fields:
\begin{eqnarray}
&&w_{\text{pair}}({\bf r},0)=w_+^0({\bf r}-{\bf r}_1)w_-^0({\bf
r}-{\bf r}_2), \label{chstatpair}
\\
&&h_{\text{pair}}({\bf r},0)=h_+^0({\bf r}-{\bf r}_1)+h_-^0({\bf
r}-{\bf r}_2), \label{hstatpair}
\\
&&{\bf u}_{\text{pair}}({\bf r},0)={\bf u}_+^0({\bf r}-{\bf
r}_1)+{\bf u}_-^0({\bf r}-{\bf r}_2). \label{ustapair}
\end{eqnarray}
For $r\rightarrow\infty$ we have ${\bf u}_+^0\sim(k_1/k_0){\bf
r}/r$ and ${\bf u}_-^0\sim -(k_2/k_0){\bf r}/r$, where $k_1$ and
$k_2$ are the inverse wavenumbers associated with the modes. Since
${\bf u}_{\text{pair}}({\bf r},0)\sim ((k_1-k_2)/k_0){\bf r}/r$ we
ensure a vanishing slope field at infinity by balancing $k_1$ and
$k_2$, i.e., $k_1=k_2=k$.

\begin{figure}
\includegraphics[width=1.0\hsize]
{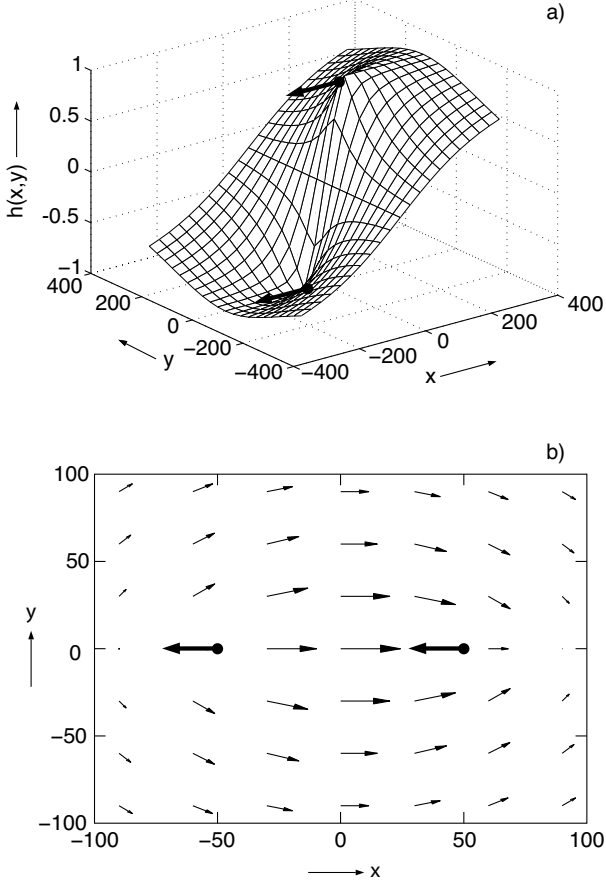}\caption{We show the height and slope field of a dipole or
pair mode in 2D. The bold arrows indicate the propagation velocity
(arbitrary units).} \label{fig13}
\end{figure}

In order to assign velocities to the modes we note that in the
vicinity of the mode position ${\bf r}_1$ the slope field is
shifted by ${\bf u}_-^0({\bf r}_1-{\bf r}_2)$ and we must,
according to the Galilei transformation (\ref{gal4}), assign the
velocity ${\bf v}_1=-\lambda{\bf u}_-^0({\bf r}_1-{\bf r}_2)$.
Likewise, the mode ${\bf u}_-^0$ is assigned velocity ${\bf
v}_2=-\lambda{\bf u}_+^0({\bf r}_2-{\bf r}_1)$. For large
separation $|{\bf r}_1-{\bf r}_2|\gg k^{-1}$ the asymptotic
expressions (\ref{linasymphu}) and (\ref{nlseasymphu}) yield ${\bf
v}_1=\lambda(k/k_0)({\bf r}_1-{\bf r}_2)/|{\bf r}_1-{\bf r}_2|$
and ${\bf v}_2=-\lambda(k/k_0)({\bf r}_2-{\bf r}_1)/|{\bf
r}_2-{\bf r}_1|$. We note that ${\bf v}_1={\bf v}_2={\bf v}$,
i.e., the individual modes propagate collectively in a direction
along the axis ${\bf r}_1-{\bf r}_2$ of the pair mode. The
propagating pair mode is thus given by
\begin{eqnarray}
&&w_{\text{pair}}({\bf r},t)=w_+^0({\bf r}-{\bf v}t-{\bf
r}_1)w_-^0({\bf r}-{\bf v}t-{\bf r}_2),~~~~~~~~ \label{chpair}
\\
&&h_{\text{pair}}({\bf r},t)=h_+^0({\bf r}-{\bf v}t-{\bf
r}_1)+h_-^0({\bf r}-{\bf v}t-{\bf r}_2),\label{hpair}
\\
&&{\bf u}_{\text{pair}}({\bf r},t)={\bf u}_+^0({\bf r}-{\bf
v}t-{\bf r}_1)+{\bf u}_-^0({\bf r}-{\bf v}t-{\bf
r}_2),\label{upair}
\\
&&{\bf v}=-\lambda {\bf u}_+^0({\bf r}_2-{\bf r}_1). \label{vpair}
\end{eqnarray}
For the purpose of our discussion and assuming a dilute gas of
growth modes it suffices to use the asymptotic forms of the
Cole-Hopf, KPZ, and Burgers fields. Introducing a core radius
$\epsilon$ of order $1/k$ in order to regularize the solution at
the origin and introducing the notation
\begin{eqnarray}
|{\bf r}|_\epsilon=\sqrt{{\bf r}^2+\epsilon^2}, \label{reg}
\end{eqnarray}
we have for the propagating pair mode fields in more detail
\begin{eqnarray}
\!\!\!\!\!&&w_{\text{pair}}({\bf r},t)\sim\exp(k|{\bf r}-{\bf
r}_1(t)|_\epsilon-k|{\bf r}-{\bf
r}_2(t)|_\epsilon),~~~~~~\label{chaympair}
\\
&&h_{\text{pair}}({\bf r},t)\sim\frac{k}{k_0}(|{\bf r}-{\bf
r}_1(t)|_\epsilon-|{\bf r}-{\bf
r}_2(t)|_\epsilon),\label{hasympair}
\\
&&{\bf u}_{\text{pair}}({\bf
r},t)\sim\frac{k}{k_0}\left(\frac{{\bf r}-{\bf r}_1(t)} {|{\bf
r}-{\bf r}_1(t)|_\epsilon}-\frac{{\bf r}-{\bf r}_2(t)}{|{\bf
r}-{\bf r}_2(t)|_\epsilon}\right),~~ \label{uasympair}
\\
&&{\bf v}=-\lambda\frac{k}{k_0}\frac{{\bf r}_2-{\bf r}_1}{|{\bf
r}_2-{\bf r}_1|_\epsilon},\label{vasympair}
\\
&&{\bf r}_{1,2}(t)={\bf v}t+{\bf r}_{1,2}. \label{pospair}
\end{eqnarray}
Denoting the separation ${\bf d}={\bf r}_2-{\bf r}_1$ and the unit
vector $\hat{\bf r}={\bf r}/r$, and expanding $h$ and ${\bf u}$
for large $r$ we obtain
\begin{eqnarray}
&&h_{\text{pair}}({\bf r})=\frac{k}{k_0}\hat{\bf r}\cdot{\bf
d},\label{hpair2}
\\
&&{\bf u}_{\text{pair}}({\bf r})={\frac{k}{k_0}\frac{1}{r}[{\bf
d}-\hat{\bf r}}(\hat{\bf r}\cdot{\bf d})].\label{upair2}
\end{eqnarray}
We note that the height field is constant at infinity, its value
depending on the direction, whereas the slope field vanishes.
Introducing polar coordinates where $\phi$ is the polar angle
between the direction $\hat{\bf r}$ and the axis ${\bf d}$ we have
\begin{eqnarray}
&&h_{\text{pair}}({\bf r})=\frac{k}{k_0}d\cos\phi, \label{hpair3}
\\
&&{\bf u}_{\text{pair}}({\bf r})=\frac{k}{k_0} \frac{1}{r}[{\bf
d}-\hat{\bf r}d\cos\phi]. \label{upair3}
\end{eqnarray}
In 1D we obtain $w_\pm^0(x)\propto\cosh^{\pm 1}kx$,
$h_\pm^0(x)\propto \pm(1/k_0)\log|\cosh kx|$,
$u_\pm^0(x)=\pm(k/k_0)\tanh kx$, and $v=-\lambda(k/k_0)$ and we
have for the pair mode the expression
\begin{eqnarray}
&&u(x,t)= \nonumber \\
&&\frac{k}{k_0}[\tanh k(x-x_1(t))-\tanh
k(x-x_2(t))],~~~~~~~\label{upair1}
\\
&&v=-\lambda\frac{k}{k_0}, \label{vpair1}
\\
&&x_{1,2}(t)=vt+x_{1,2}, \label{pos1dpair}
\end{eqnarray}
in accordance with Eq. (\ref{upair}) and the discussion in Refs.
\cite{Fogedby03b}; the 1D pair mode is depicted in
Fig.~\ref{fig6}. In Fig.~\ref{fig13} we have depicted the height
and slope fields in 2D.

The propagation of the pair mode is in the direction of the axis
connecting the two centers ${\bf r}_1$ and  ${\bf r}_1$. During
the passage of a pair with velocity $v=\lambda k/k_0$ it follows
from Eqs. (\ref{hpair3}) and (\ref{upair3}) that the local height
and slope fields change by $\delta h=2kd/k_0$ and $\delta
u=2k/k_0$, respectively. The passage time of the pair is $\delta
t=d/v=dk_0/\lambda k$ and we obtain for the growth velocity
$\delta h/\delta t=2\lambda(k/k_0)^2=(\lambda/2)(\delta u)^2$, in
accordance with the averaged KPZ equation in a stationary state,
$\langle\delta h/\delta t\rangle=(\lambda/2){\bf u}^2$.

\begin{figure}
\includegraphics[width=1.0\hsize]
{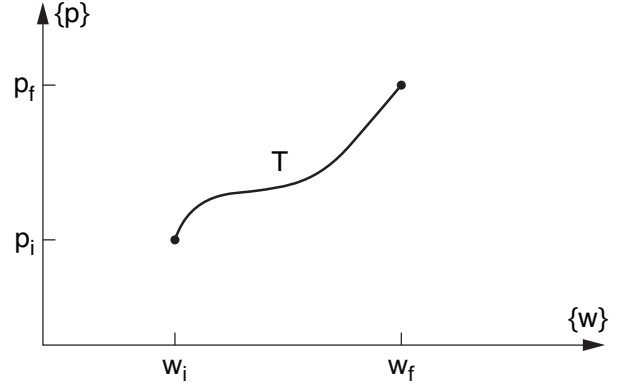}\caption{Phase space plot of a dipole or pair orbit from
an initial configuration $w_{\text{i}}$ to a final configuration
$w_{\text{f}}$ in transition time $T$. The noise field $p$ is a
slaved variable going from $p_{\text{i}}$ to $p_{\text{f}}$.}
\label{fig14}
\end{figure}

The dynamics of the pair mode is inferred from Eqs. (\ref{chS1}),
(\ref{chE2}), and (\ref{chP2}). Inserting Eq. (\ref{nlsedimless})
a scaling argument yields
\begin{eqnarray}
S_{\text{pair}}^{\text{CH}} =
\frac{1}{2}(k_0\nu)^2T\left(\frac{k}{k_0}\right)^4 a_d^4
k^{-d}\int_0^\infty d^d\xi~f(\xi)^4, \label{chpairS}
\end{eqnarray}
for the action of a pair propagating in time $T$. Note that only
the decaying bound state component carries action. In the
dynamical phase space language this scenario corresponds to a
pair-orbit from an initial configuration $w_{\text{i}}({\bf
r},0)=w_+^0({\bf r}-{\bf r}_1)w_-^0({\bf r}-{\bf r}_2)$ with
initial noise field $p_{\text{i}}({\bf r},0)=\nu w_-^0({\bf
r}-{\bf r}_2)$ to a final configuration $w_{\text{f}}({\bf
r},T)=w_+^0({\bf r}-{\bf v}T-{\bf r}_1)w_-^0({\bf r}-{\bf v}T-{\bf
r}_2)$ and $p_{\text{f}}({\bf r},T)=\nu w_-^0({\bf r}-{\bf
v}T-{\bf r}_2)$. In Fig.~\ref{fig14} we have in a $(w,p)$ phase
space plot sketched the specific orbit from
$(w_{\text{i}},p_{\text{i}})$ to $(w_{\text{f}},p_{\text{f}})$.
Choosing the centers ${\bf r}_1$ and ${\bf r}_2$ and the amplitude
$k$ we thus have a whole class of orbits corresponding to kinetic
transitions from $({\bf r}_1,{\bf r}_2,k)$ at time $t=0$ to $({\bf
r}_1+{\bf v}T,{\bf r}_2+{\bf v}T,k)$ at time $T$; note that the
magnitude of the velocity is $v=\lambda k/k_0$, whereas its
direction is given by ${\bf d}={\bf r}_1-{\bf r}_2$.

For the rest energy and mass of the pair mode we have, inserting
Eq. (\ref{nlsedimless}) in Eqs. (\ref{chE2}), (\ref{chP2}),
(\ref{chm}), and (\ref{chrestE}),
\begin{eqnarray}
&&E_{0,\text{pair}} =
-\frac{1}{2}(k_0\nu)^2T\left(\frac{k}{k_0}\right)^4 a_d^4
k^{-d}\int d^d\xi~f(\xi)^4,~~~~~~~ \label{chpairE0}
\\
&&m_{\text{pair}}=\frac{1}{2}\left(\frac{k}{k_0}\right)^2a_d^2k^{-d}
\int d^d\xi~f(\xi)^2, \label{chpairm}
\end{eqnarray}
and we obtain ~for the energy and momentum
\begin{eqnarray}
&&E_{\text{pair}}=E_{0,\text{pair}}+\frac{1}{2}m_{\text{pair}}{\bf
v}^2,\label{chpairE}
\\
&&{\bm\Pi}_{\text{pair}}=m_{\text{pair}}{\bf v}, \label{chpairP}
\\
&&{\bf v}=-\lambda\frac{k}{k_0}{\bf d}. \label{chpairv}
\end{eqnarray}
The pair mode satisfying the boundary conditions of an
asymptotically flat interface suggests an independent particle
picture of the growth morphology in terms of a dilute gas of pair
modes. The pair modes have masses scaling with the amplitude
according to $m_{\text{pair}}\propto k^{2-d}$. In 2D the mass is
independent of $k$, in 1D the mass grows linearly with $k$, for
$d>2$ the mass vanishes for large $k$. The rest energy
$E_{0,\text{pair}}\propto k^{4-d}$, i.e., $E_{0,\text{pair}}$ is
independent of $k$ for $d=4$. For $d<4$, $E_{0,\text{pair}}$ grows
with $k$, for $d>4$, $E_{0,\text{pair}}$ vanishes for large $k$.
Finally, the velocity scales linearly with $k$.
\subsection{Dynamical network}
The generalization to a network of propagating modes is
straightforward. At time $t=0$ we assign a set of growing and
decaying static modes at the positions ${\bf r}_i^0$,
$i=1,\cdots$, and obtain the Cole-Hopf field at time $t=0$,
\begin{eqnarray}
w({\bf r},0)=\prod_iw_i^0({\bf r}-{\bf r}_i^0). \label{chstatnet}
\end{eqnarray}
It is here convenient to use a charge language for the amplitudes
$k_i$ pertaining to the $i$-th mode. For $k_i>0$, a positive
charge, $w_i^0({\bf r})$ is a growing solution of the linear
equation (\ref{chlin}); for $k_i<0$, a negative charge,
$w_i^0({\bf r})$ is a decaying bound state solution of the NLSE
(\ref{chnlse}). For large $r$ we have $w_i^0({\bf r}-{\bf
r}_i^0)\sim \exp(k_ir)$, i.e., $w({\bf r},0)\sim\exp(r\sum_i k_i)$
and in order to ensure an asymptotically flat interface $h({\bf
r})=(1/k_0)\log w({\bf r})$ we impose the charge neutrality
condition
\begin{eqnarray}
\sum_i k_i=0.  \label{neut}
\end{eqnarray}
The corresponding initial height and slope field are then  given
by
\begin{eqnarray}
&&h({\bf r},0)=\frac{1}{k_0}\sum_i\log w_i^0({\bf r}-{\bf
r}_i^0),\label{hstatnet}
\\
&&{\bf u}({\bf
r},0)=\frac{1}{k_0}\sum_i\frac{{\bm\nabla}w_i^0({\bf r}-{\bf
r}_i^0)}{w_i^0({\bf r}-{\bf r}_i^0)}.\label{ustatnet}
\end{eqnarray}
In order to assign velocities to the modes we proceed as in the
case of the pair mode. At the position ${\bf r}^0_\ell$ of the
$\ell$-th mode the slope field is shifted by ${\bf u}({\bf
r}_\ell^0,0)=(1/k_0)\sum_i{\bm\nabla}w_i^0({\bf r}_\ell^0-{\bf
r}_i^0)/w_i^0({\bf r}_\ell^0-{\bf r}_i^0)$, corresponding to a
Galilei boost to velocity ${\bf v}_\ell=\lambda{\bf u}({\bf
r}_\ell^0,0)$. Since, unlike the case of a pair, the modes move
relative to one another the network will converge towards a self
consistent state. We thus obtain the self consistent dynamical
network
\begin{eqnarray}
&&w({\bf r},t)=\prod_iw_i^0({\bf r}-{\bf r}_i(t)), \label{chnet}
\\
&&{\bf r}_i(t)=\int_0^t{\bf v}_i(t')dt'+{\bf r}_i^0,
\label{posnet}
\\
&&{\bf v}_i(t)=-2\nu\sum_{\ell\neq i}\frac{{\bm\nabla}w_i^0({\bf
r}_i(t)-{\bf r}_\ell(t))}{w_i^0({\bf r}_i(t)-{\bf r}_\ell(t))},
\label{velnet}
\\
&&\sum_ik_i=0. \label{neut2}
\end{eqnarray}
The interpretation of the growth morphology or pattern formation
represented by Eq. (\ref{chnet}-\ref{neut2}) is straightforward.
In the weak noise WKB representation the growing interface is
described by a gas of growth modes with negative and positive
charges. The asymptotic flatness condition is ensured by imposing
charge neutrality as expressed by Eq. (\ref{neut2}). The dynamics
of the network is constrained by the assignment of velocities to
the modes, where according to Eq. (\ref{velnet}) the velocity of a
particular mode depends on the position and charges of the other
modes. The connectivity and continuity of the network thus defines
the temporal evolution.

For the present purposes it is sufficient to consider a dilute
network and use the asymptotic form $w_i^0({\bf
r})\sim\exp(k_ir)$. We then obtain the height field, slope field,
and assigned velocities
\begin{eqnarray}
&&h({\bf r},t)=\frac{1}{k_0}\sum_ik_i|{\bf r}-{\bf
r}_i(t)|_\epsilon,\label{hasymnet}
\\
&&{\bf u}({\bf r},t)=\frac{1}{k_0}\sum_ik_i\frac{{\bf r}-{\bf
r}_i(t)}{|{\bf r}-{\bf r}_i(t)|_\epsilon},\label{uasymnet}
\\
&&{\bf v}_i(t)=-2\nu\sum_{\ell\neq i}k_\ell\frac{{\bf r}_i(t)-{\bf
r}_\ell(t)}{|{\bf r}_i(t)-{\bf r}_\ell(t)|_\epsilon},
\label{velasymnet}
\end{eqnarray}
where we have used the short distance or UV regularization given
by Eq. (\ref{reg}).

Since we initialize the network at rest, it will pass thorough a
transient period where the velocities adjust to constant values as
the modes recedes from one another. From Eq. (\ref{velasymnet}) we
obtain by differentiation $d{\bf v}_i/dt=-2\nu\sum_{\ell\neq
i}[({\bf v}_i-{\bf v}_\ell)/|{\bf r}_i-{\bf
r}_\ell|_\epsilon-({\bf r}_i-{\bf r}_\ell)\cdot({\bf v}_i-{\bf
v}_\ell)({\bf r}_i-{\bf r}_\ell)/|{\bf r}_i-{\bf
r}_\ell|_\epsilon^3]$ which, assuming ${\bf v}_i-{\bf v}_\ell$ to
be bounded, vanishes for $|{\bf r}_i-{\bf
r}_\ell|\rightarrow\infty$. We infer that at intermediate times
longer than the transient time the velocities attain constant
values. Using ${\bf r}_i(t)={\bf v}_it+{\bf r}_i^0$ we thus obtain
from Eq. (\ref{velasymnet}) a self consistent equation for the
velocities in the stationary state
\begin{eqnarray}
{\bf v}_i=-2\nu\sum_{\ell\neq i}k_\ell\frac{{\bf v}_i-{\bf
v}_\ell}{|{\bf v}_i-{\bf v}_\ell|_\epsilon}. \label{vstatnet}
\end{eqnarray}

At fixed time for ${\bf r}$ large expanding Eq. (\ref{hasymnet})
we obtain
\begin{eqnarray}
h({\bf r},t)\sim -\frac{1}{k_0}\frac{{\bf r}}{r}\sum_ik_i{\bf
r}_i(t), \label{hrlarge}
\end{eqnarray}
where we have used the neutrality condition (\ref{neut2}), i.e.,
an asymptotically flat interface. $\sum_ik_i{\bf r}_i(t)$ defines
a center of mass position ${\bf R}(t)=\sum_ik_i{\bf r}_i(t)$ for a
dilute cluster of modes. Introducing the polar angle $\phi$
between the direction ${\bf r}/r$ and ${\bf R}(t)$ we have, in
analogy to Eqs. (\ref{hpair3}) and (\ref{upair3}) in the case of
the pair mode, $h({\bf r},t)=-(1/k_0)|{\bf R}(t)|\cos\phi$, and
the height offset depends on the direction. As the cluster of
modes propagate across the system the height field changes by
$2|{\bf R}(t)|/k_0$. Likewise, for the slope field ${\bf u}$
expanding Eq. (\ref{uasymnet}) for large ${\bf r}$ we have ${\bf
u}({\bf r},t)\sim({\bf r}/r)\sum_ik_i=0$, i.e., asymptotically a
flat interface. We also note that ${\bf R}(t)=\sum_ik_i({\bf
r}_i^0+{\bf v}_it)=\sum_ik_i({\bf r}_i^0+{\bf v}_it)=\sum_ik_i{\bf
r}_i^0$, where we have used $\sum_ik_i{\bf v}_i=0$, following from
Eq. (\ref{vstatnet}), i.e., ${\bf R}(t)=\sum_ik_i{\bf r}_i^0$ is
independent of time in the stationary state. Likewise, at fixed
position for $t$ large we have by expanding Eq. (\ref{hasymnet}),
$h({\bf r},t)=(t/k_0)\sum_ik_iv_i$, and we infer the constant
growth velocity
\begin{eqnarray}
v=\frac{dh}{dt}=\frac{1}{k_0}\sum_ik_iv_i. \label{grv}
\end{eqnarray}

The relationship between the imposed drift $F$ in the KPZ equation
(\ref{kpz}) and the charge assignment to the growth modes,
$\{k_i\}$, is given by
\begin{eqnarray}
F=\frac{2\nu^2}{\lambda}\left(\sum_ik_i\right)^2.\label{relfk}
\end{eqnarray}
In order to demonstrate this identity we consider the Cole-Hopf
field equation (\ref{chdet}) in the asymptotic regions where the
noise field vanishes, $\partial w/\partial t=\nu{\bm\nabla}^2w-\nu
k^2w$, $k^2=\lambda F/2\nu^2$. From the growth mode ansatz
$w=\prod_i\exp k_i|{\bf r}-{\bf r}_i(t)|$ we readily obtain
$(\partial w/\partial t)/w=-\sum_ik_i{\bf v}_i\cdot({\bf r}-{\bf
r}_i(t))/|{\bf r}-{\bf r}_i(t)|$ and
${\bm\nabla}^2w/w=\sum_{ij}k_ik_j({\bf r}-{\bf r}_i(t))\cdot({\bf
r}-{\bf r}_j(t))/|{\bf r}-{\bf r}_i(t)||{\bf r}-{\bf r}_j(t)|$.
Inserting the mode velocity from Eq. (\ref{velasymnet}) and
symmetrizing we have for large ${\bf r}$ the identity
(\ref{relfk}). The drift $F$ is thus given by the total charge of
the growth modes. A neutral system with vanishing slope at
infinity corresponds to $F=0$.

In addition to the dynamical velocity constraint imposed by the
continuity and connectivity of the network, the canonical structure
of the WKB weak noise scheme impart dynamical attributes to the
network. As far as the dynamics is concerned only modes with
negative charge on the noise manifolds $p_i=\nu w_i$, corresponding
to the bound state solution of the NLSE, contribute. According to
Eq. (\ref{chS1}) we obtain for the total action of the network
\begin{eqnarray}
S_{\text{net}}=\sum_{k_i<0}S_i, \label{actnet}
\end{eqnarray}
where the action of the i-th mode is
\begin{eqnarray}
S_i=\frac{1}{2}(k_0\nu)^2T\int d^dx~w_i^0(r)^4.\label{actmode}
\end{eqnarray}
For a dilute network using the asymptotic expression
(\ref{chpairS}) we obtain accordingly
\begin{eqnarray}
S_i = \frac{1}{2}(k_0\nu)^2T\left(\frac{k_i}{k_0}\right)^4 a_d^4
k^{-d}\int_0^\infty d^d\xi~f(\xi)^4. \label{actmode2}
\end{eqnarray}

At time $t=0$ we assign an initial state $w_i({\bf r},0)$ by
choosing a set of positions ${\bf r}_i^0$ and associated charges
$k_i$. The expressions (\ref{velasymnet}) and (\ref{posnet})
subsequently determine the appropriate propagation velocities and
time dependent positions. The network develops dynamically
defining a particular kinetic pathway.  In order to obtain a
specific path from an initial state $w_{\text{i}}$ to a final
state $w_{\text{f}}$ traversed in time $T$ we must choose a
specific set of positions and charges. The action associated with
the path is then given by Eqs. (\ref{actnet}-\ref{actmode2}),
yielding the Arrhenius factor $P\propto\exp[-S/\Delta]$ for the
corresponding transition probability.

\begin{figure}
\includegraphics[width=1.0\hsize]
{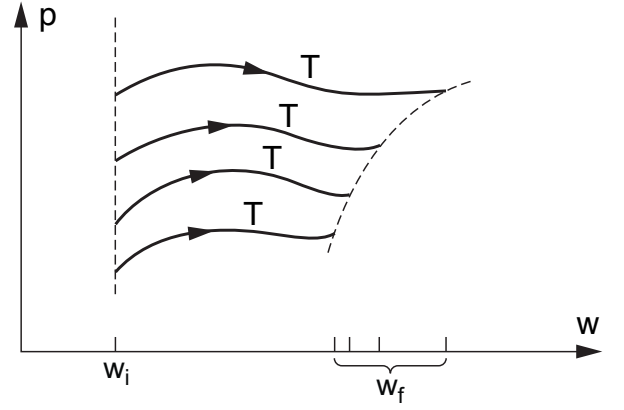}\caption{Phase space plot of a set of orbits with the same
transition time $T$ terminating in different final configurations
$w_{\text{f}}$.} \label{fig15}
\end{figure}

Associated with the initial configuration $w_{\text{i}}$ at time
$t=0$ is the initial noise field
\begin{eqnarray}
p_{\text{i}}({\bf r},0)\propto\prod_{k_i<0}w_i^0({\bf r}-{\bf
r}_i^0).\label{initp}
\end{eqnarray}
Assigning velocities the noise field propagates according to
\begin{eqnarray}
p({\bf r},t)\propto\prod_{k_i<0}w_i^0({\bf r}-{\bf v}_it-{\bf
r}_i^0), \label{evolp}
\end{eqnarray}
to the final noise configuration at time $T$. For a fixed
transition time $T$ and a given initial $w$-configuration
different assignments of $k_i$ and ${\bf r}_i^0$ yield different
noise field and, correspondingly, different final $(w,p)$
configurations. The scenario is depicted schematically in
Fig.~\ref{fig15}

The energy and momentum of a network is inferred from Eqs.
(\ref{chpairE0}-\ref{chpairv})
\begin{eqnarray}
&&E_{\text{net}}=E_{\text{net},0}+\frac{1}{2}\sum_im_iv_i^2,
\label{enet}
\\
&&E_{\text{net},0}= \nonumber
\\
&& -\frac{1}{2}(k_0\nu)^2
\sum_{k_i<0}\left(\frac{k_i}{k_0}\right)^4 a_d^4 k_i^{-d}\int
d^d\xi~f(\xi)^4,~~~~~~~\label{E0net}
\\
&&m_i=\frac{1}{2}\left(\frac{k_i}{k_0}\right)^2 a_d^2 k_i^{-d}\int
d^d\xi~f(\xi)^2, \label{mi}
\\
&&{\bm\Pi}_{\text{net}}=\sum_{k_i<0}m_i{\bf v}_i, \label{Pnet}
\end{eqnarray}
where in the stationary state ${\bf v}_i$ is given by Eq.
(\ref{vstatnet}).
%
\subsection{Field theory}
In a qualitative sense the Cole-Hopf field $w$ and the associated
noise field $p$ governed by Eqs. (\ref{cheq1}) and (\ref{cheq2})
representing the KPZ equation in the weak noise limit plays the
role of bare fields, whereas the propagating noiseless and noisy
modes $w_i^0$ together with the associated noise fields connected
in a dynamical network constitute the renormalized fields. The
strong coupling features of the problem are thus represented by
the network. It is instructive to represent this insight in terms
of a field theory for the network. Below we sketch aspects of such
a field theory.

We consider a dilute distribution of modes or monopoles with
charges $k_i$ at positions $r_i$ and introduce the density $n$ and
charge density $\rho$ according to
\begin{eqnarray}
&&n({\bf r},t)=\sum_i\delta^d({\bf r}-{\bf r}_i(t)), \label{den}
\\
&&\rho({\bf r},t)=\sum_i k_i\delta^d({\bf r}-{\bf r}_i(t));
\label{chden}
\end{eqnarray}
we note that the neutrality condition $\sum_ik_i=0$ is equivalent
to $\int d^dx\rho({\bf r},t)=0$. For the asymptotic height and
slope fields (\ref{hasymnet}) and (\ref{uasymnet}) we obtain
\begin{eqnarray}
&&h({\bf r},t)=\frac{1}{k_0}\int d^dx'~\rho({\bf r}',t)|{\bf
r}-{\bf r}'|_\epsilon, \label{hfield} \\
&&{\bf u}({\bf r},t)=\frac{1}{k_0}\int d^dx'~\rho({\bf
r}',t)\frac{{\bf r}-{\bf r}'}{|{\bf r}-{\bf r}'|_\epsilon}.
\label{ufield}
\end{eqnarray}
Introducing the velocity field
\begin{eqnarray}
{\bf v}({\bf r},t)=\sum_i{\bf v}_i(t) \delta^d({\bf r} -{\bf
r}_i(t)), \label{vfield}
\end{eqnarray}
we express the velocity condition (\ref{velasymnet}) in the form
\begin{eqnarray}
{\bf v}({\bf r},t)=-2\nu n({\bf r},t){\bf u}({\bf r},t).
\label{vfield2}
\end{eqnarray}
Using ${\bm\nabla}\cdot({\bf r}/r)=(d-1)/r$ and introducing the
scalar field or potential $\phi$
\begin{eqnarray}
\phi={\bm\nabla}\cdot{\bf u}=\nabla^2h, \label{potdef}
\end{eqnarray}
we have from Eq. (\ref{vfield})
\begin{eqnarray}
\phi({\bf r},t)=\frac{d-1}{k_0}\int d^dx'~\frac{\rho({\bf
r}',t)}{|{\bf r}-{\bf r}'|_\epsilon}, \label{pot}
\end{eqnarray}
i.e., $\phi$ is analogous to the electric potential arising from a
charge distribution $\rho({\bf r},t)$ \cite{Landau60}. It also
follow from Eq (\ref{pot}) that $\phi$ satisfies the fractional
Poisson equation
\begin{eqnarray}
\nabla^{d-1}\propto\frac{d-1}{k_0}\rho, \label{upoisson}
\end{eqnarray}
where $\nabla^{d-1}$ is the Fourier transform of $k^{d-1}$; note
that in $d=3$ Eq. (\ref{upoisson}) becomes the usual Poisson
equation in electrostatics \cite{Landau60}.

The charge distribution $\rho$ yields the potential $\phi$ either
as a solution of Eq. (\ref{upoisson}) or in terms of the
integrated form in Eq. (\ref{pot}). Inserting Eq. (\ref{potdef})
we also have
\begin{eqnarray}
&&\nabla^{d+1}h\propto\frac{d-1}{k_0}\rho, \label{hpoisson}
\\
&&\nabla^{d-1}{\bm\nabla}\cdot{\bf u}\propto\frac{d-1}{k_0}\rho.
\label{upoisson2}
\end{eqnarray}
The equation of motion for the charges is given by Eq.
(\ref{vfield2}) which determine the velocity field ${\bf v}$ in
terms of the slope field ${\bf u}$. It is instructive to compare
Eq. (\ref{vfield}) to the Lorentz equation for the motion of
charges $e$ with density $n$ in an electric field ${\bf E}$,
$md{\bf v}/dt=en{\bf E}$. Including a damping term $-\gamma{\bf
v}$ and considering the overdamped case we obtain ${\bf
v}=-(e/\gamma)n{\bf E}$ which has the same form as Eq.
(\ref{vfield2}) with the slope field playing the role of an
electric field.

\begin{figure}
\includegraphics[width=1.0\hsize]
{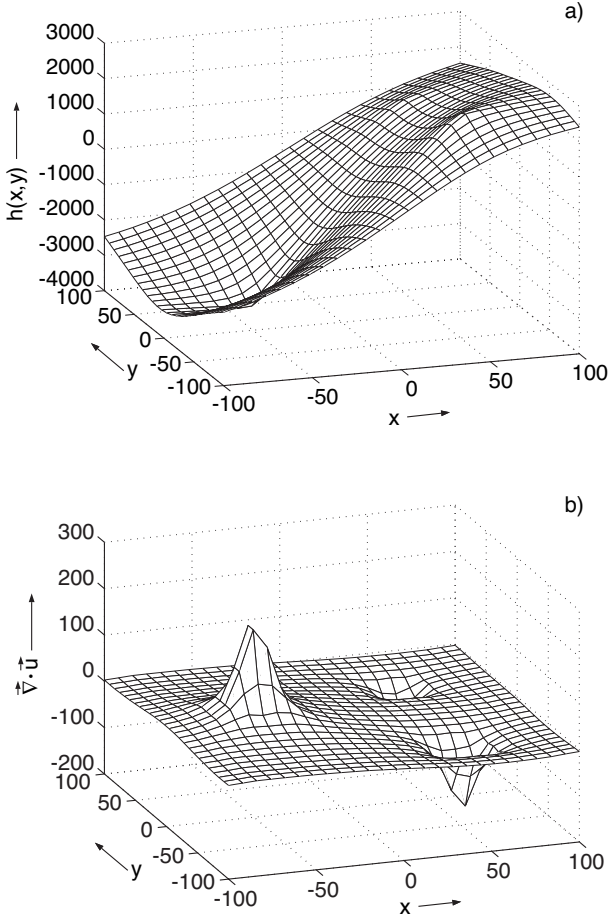}\caption{We depict a neutral three-mode configuration in
the height field and the associated divergence of the slope field
(arbitrary units).} \label{fig16}
\end{figure}

Finally, introducing the continuity equations expressing number
and charge conservation
\begin{eqnarray}
&&\frac{\partial n}{\partial t}+{\bm\nabla}\cdot{\bf v}=0,
\label{con1}
\\
&&\frac{\partial\rho}{\partial t}+{\bm\nabla}\cdot{\bf j}=0,
\label{con2}
\end{eqnarray}
where the current density is
\begin{eqnarray}
{\bf j}({\bf r},t)=\sum_ik_i{\bf v}_i\delta^d({\bf r}-{\bf
r}_i(t)), \label{curden}
\end{eqnarray}
we have the field equation (\ref{upoisson2}) for ${\bf u}$ and the
Lorentz equation (\ref{vfield2}) for the particle dynamics. In
Fig.~\ref{fig16} we depict a neutral three-mode configuration in
the height field and the associated divergence of the slope field.
${\bm\nabla}\cdot{\bf u}$ clearly shows the two negative charges
and the single positive charge constituting the morphology.
%
\subsection{Linear diffusive modes}
Generalizing the discussion in the case of the 1D noisy Burgers
equation to higher dimension, it is clear that in addition to the
network of growth modes there are also superimposed linear
diffusive modes. Here we summarize aspects of the linear mode
spectrum.
\subsubsection{The linear case}
In the linear Edwards-Wilkinson case the weak noise field
equations (\ref{bureq1}) and (\ref{bureq2}) take the form
\begin{eqnarray}
&&\frac{\partial{\bf u}}{\partial t}=\nu\nabla^2{\bf
u}-{\bm\nabla}({\bm\nabla}\cdot{\bf p}), \label{linbur1}
\\
&&\frac{\partial{\bf p}}{\partial t}=-\nu\nabla^2{\bf u}.
\label{linbur2}
\end{eqnarray}
Since ${\bf u} (={\bm\nabla}h)$ is longitudinal and since only the
longitudinal component of ${\bf p}$ couple to ${\bf u}$, Eqs.
(\ref{linbur1}) and (\ref{linbur2}) for the wavenumber components
${\bf u}_{\bf k}$ and ${\bf p}_{\bf k}$ correspond to the
overdamped oscillator case discussed in appendix \ref{app}. For an
orbit from ${\bf u}^{\text{i}}_{\bf k}$ at time $t=0$ to ${\bf
u}^{\text{f}}_{\bf k}$ at $t=T$, with ${\bf p}_{\bf k}$ as slaved
variable, we have
\begin{eqnarray}
&&{\bf u}_{\bf k}(t)=\frac{{\bf u}^f_{\bf k}\sinh\omega_{\bf
k}t+{\bf u}^i_{\bf k}\sinh\omega_{\bf k}(T-t)}{\sinh\omega_{\bf
k}T}, \label{sollinbur1}
\\
&&{\bf p}_{\bf k}(t)=\nu e^{\omega_{\bf k}t}\frac{{\bf u}^f_{\bf
k}-{\bf u}^i_{\bf k} e^{-\omega_{\bf k}T}}{\sinh\omega_{\bf
k}T},\label{sollinbur2}
\end{eqnarray}
with the diffusive mode frequency
\begin{eqnarray}
\omega_{\bf k}=\nu {\bf k}^2. \label{damp}
\end{eqnarray}
The spectrum is exhausted by linear diffusive modes with gapless
dispersion given by Eq. (\ref{damp}). Following Ref.
\cite{Fogedby03b} the action and transition probabilities are
given by
\begin{eqnarray}
&&S=\nu\int\frac{d^dk}{(2\pi)^d}~\frac{|{\bf u}_{\bf k}^f-{\bf
u}^i_{\bf k}\exp(-\omega_{\bf k}T)|^2}{1-\exp(-2\omega_{\bf k}T)},
\label{actlin}
\\
&&P({\bf u}_{\bf k}^f,{\bf u}^i_{\bf
k},T)\propto\exp\left[-\frac{S}{\Delta}\right]. \label{dislin}
\end{eqnarray}
In the limit $T\rightarrow\infty$ we obtain the stationary
distribution
\begin{eqnarray}
P_0({\bf u}_{\bf k})\propto\exp\left[-\frac{\nu}{\Delta}
\int\frac{d^dk}{(2\pi)^d}~|{\bf u}_{\bf k}|^2\right],
\label{stalin}
\end{eqnarray}
yielding the Boltzmann distribution (\ref{stakpz}); for more
comments on the linear case see the discussion in
Sec.~\ref{kpzeq}.
\subsubsection{The nonlinear growth case}
In the nonlinear KPZ case the growth morphology or pattern
formation is given by a dynamical network of propagating localized
growth modes. Superimposed on the dynamical network is a spectrum
of extended linear diffusive modes. In order to implement the
boundary condition of an asymptotically flat interface it is most
convenient to conduct the discussion in terms of the slope field
${\bf u}$.

Considering the field equations (\ref{bureq1}) and (\ref{bureq2})
we set ${\bf u}={\bf u}_{\text{net}}+\delta{\bf u}$ and ${\bf
p}={\bf p}_{\text{net}}+\delta{\bf p}$. Here ${\bf
u}_{\text{net}}$ is given by (\ref{uasymnet}), i.e.,
\begin{eqnarray}
{\bf u}_{\text{net}}({\bf r},t)=\frac{1}{k_0}\sum_ik_i\frac{{\bf
r}-{\bf r}_i(t)}{|{\bf r}-{\bf r}_i(t)|_\epsilon}.\label{unet}
\end{eqnarray}
For the associated noise field ${\bf p}_{\text{net}}$ we obtain
from Eqs. (\ref{kpzch}), (\ref{nlseasymp}), and (\ref{burkpz}) for
a single mode $\tilde p=k_0pw=k_0\nu w_-^2\propto k_0\nu
r^{1-d}\exp(-2kr)$, i.e., ${\bm\nabla}\cdot{\bf p}=-k_0\nu
r^{1-d}\exp(-2kr)$. Setting ${\bf p}=({\bf r}/r)g(r)$ we have
$dg/dr+((d-1)/r)g=-k_0\nu r^{1-d}\exp(-2kr)$ with asymptotic
solution $g(r)\sim (k_0\nu/2k)r^{1-d}\exp(-2kr)$. Generalizing to
a propagating network we have
\begin{eqnarray}
{\bf p}_{\text{net}}({\bf r},t)=\frac{\nu
k_0}{2}\sum_{k_i<0}\frac{{\bf r}-{\bf r}_i(t)}{|k_i||{\bf r}-{\bf
r}_i(t)|^d_\epsilon}e^{-2|k_i||{\bf r}-{\bf
r}_i(t)|}.~~\label{pnet}
\end{eqnarray}
We note that the noise field is localized in the vicinity of the
modes with negative charge, i.e., the modes carrying dynamics. The
noise field falls off exponentially with a range given by
$|k_i|^{-1}$. Confining our discussion to the regions between the
growth modes and noting that $p_{\text{net}}\sim 0$ and $\nabla{\bf
u_{\text{net}}}\sim 0$ we obtain linear equations for $\delta{\bf
u}$ and $\delta{\bf p}$,
\begin{eqnarray}
&&\frac{\partial\delta{\bf u}}{\partial t}=\nu\nabla^2\delta{\bf
u}+\lambda({\bf u}_{\text{net}}\cdot{\bm\nabla})\delta{\bf
u}+{\bm\nabla}({\bm\nabla}\cdot\delta{\bf p}),~~~~ \label{linu}
\\
&&\frac{\partial\delta{\bf p}}{\partial t}=-\nu\nabla^2\delta{\bf
p}+\lambda({\bf u}_{\text{net}}\cdot{\bm\nabla})\delta{\bf p}.
\label{linp}
\end{eqnarray}
Assuming that ${\bf u}_{\text{net}}\sim\text{const.}$ and setting
$\delta{\bf u}, \delta{\bf p}\sim\exp(-Et)\exp(i{\bf k}{\bf r})$
it follows that
\begin{eqnarray}
&&\delta{\bf u}=({\bf a}e^{-(\nu k^2-i\lambda{\bf
u}_{\text{net}}\cdot{\bf k})t} + {\bf b}e^{(\nu k^2+i\lambda{\bf
u}_{\text{net}}\cdot{\bf k})t})e^{i{\bf k}{\bf r}},~~~~~~~
\label{linu2}
\\
&&\delta{\bf p}={\bf c}e^{(\nu k^2+i\lambda{\bf
u}_{\text{net}}\cdot{\bf k})t}e^{i{\bf k}{\bf r}}. \label{linp2}
\end{eqnarray}
For $\lambda=0$ we recover the linear case; in the nonlinear case,
incorporating the time dependent term $i\lambda{\bf
u}_{\text{net}}\cdot{\bf k}t$ in the plane wave phase $i{\bf
k}{\bf r}$, we note that the diffusive modes undergoes a mode
transmutation to damped and growing propagating modes with phase
velocity $\lambda{\bf u}_{\text{net}}\cdot{\bf k}$.
\section{\label{scaling} Stochastic interpretation}
The weak noise scheme gives access to the transition probability
$P(h_{\text{i}}\rightarrow h_{\text{f}},T)$ from an initial height
profile $h_{\text{i}}$ to a final profile $h_{\text{f}}$ in time
$T$, where $P$ is given in terms of the action
$S(h_{\text{i}}\rightarrow h_{\text{f}},T)$,
$P\propto\exp[-S/\Delta]$. We note that the scheme only yields
what corresponds to the Arrhenius factor $\exp[-S/\Delta]$; the
prefactor $\Gamma(T)$, $P(T)=\Gamma(T)\exp[-S(T)/\Delta]$, is to
leading order in $\Delta$ determined by the normalization
condition $\int\prod_{\bf r}dh_{\text{f}}P(h_{\text{i}}\rightarrow
h_{\text{f}},T)=1$, i.e., $\Gamma^{-1}(T)=\int\prod_{\bf
r}dh_{\text{f}}P(h_{\text{i}}\rightarrow h_{\text{f}},T)$, see
also Refs. \cite{Fogedby03b,Fogedby04b}.
\subsection{Kinetic transitions}
The prescription is in principle straightforward. An initial
configuration $h_{\text{i}}$ at time $t=0$ is modelled or
approximated by a dilute gas of growth modes forming the network
characterized by their charges $\{k_i\}$ and positions $\{{\bf
r}_i^0\}$ plus a spectrum of diffusive modes $\{\delta
u_k^{\text{i}},\delta p_k^{\text{i}}\}$. The dynamical
configuration evolves in time according to the field equations.
The velocities of the growth modes are assigned according to Eqs.
(\ref{velnet}) and (\ref{velasymnet}); notice that the noise field
associated with the negatively charged modes develops in time
according to Eq. (\ref{pnet}). At time $t=T$ the profile $h$ has
evolved to the final profile $h_{\text{f}}$, corresponding to the
network configuration $\{k_i,{\bf r}_i(T)\}$ and the final
diffusive mode configuration $\{\delta u_k^{\text{f}},\delta
p_k^{\text{f}}\}$. This time evolution corresponds to a specific
kinetic pathway for a growing interface. The action associated
with the transition is composed of a network part $S_{\text{net}}$
and a diffusive mode part $S_{\text{diff}}$. $S_{\text{net}}$ is
given by Eqs. (\ref{actnet}), (\ref{actmode}), and
(\ref{actmode2}) and thus only depends on $\{k_i\}$ for the
negatively charged bound states or monopoles. In the linear case
$S_{\text{diff}}$ is given by Eq. (\ref{actlin}); we note that
$S_{\text{diff}}$ depends on the initial and final diffusive mode
amplitudes $\delta u_k^{\text{i}}$ and $\delta u_k^{\text{f}}$. At
long times $S_{\text{net}}$ grows linearly with $T$, like in the
case of random walk discussed in appendix \ref{app}, whereas
$S_{\text{diff}}$ approaches the stationary form $\nu\int
d^dk/(2\pi)^d|{\bf u}_k^{\text{f}}|^2$, see Eq. (\ref{stalin}), as
in the case of the overdamped oscillator discussed in appendix
\ref{app}.
\subsection{Anomalous diffusion and scaling in the dipole sector}
Leaving aside the issue of the linear diffusive modes, the network
representation is based on an assumption of a dilute gas of growth
modes or charged monopoles. In the course of time the modes will
in general collide and coalesce and the dilute gas approximation
ceases to be valid. Here we consider the class of network
configurations composed of a dilute gas of pair modes or dipoles.
A single dipole satisfies the boundary condition of vanishing
slope field. Consequently, the dipoles move independently.

A single dipole or pair mode with charges $k$ and $-k$ propagate
according to Eq. (\ref{chpairv}) with velocity
\begin{eqnarray}
v=\lambda\frac{k}{k_0}, \label{vdip}
\end{eqnarray}
and carries according to Eq. (\ref{chpairS}) the action
\begin{eqnarray}
S=Tk^{4-d}\left(\frac{\nu}{k_0}\right)^2A, \label{sdip}
\end{eqnarray}
where $A=a_d^4\int_0^\infty d^d\xi~f(\xi)$ only depends on
dimension. During time $T$ the center of mass of the dipole
propagates the distance $L=vT$ and eliminating the charge $k$ we
obtain the action
\begin{eqnarray}
S=\frac{L^{4-d}}{T^{3-d}}B,\label{sdip2}
\end{eqnarray}
where $B=\lambda^{-2}\nu^d 2^{d-2}A$.

The form of Eq. (\ref{sdip2}) allows an interpretation of the
ballistic motion of the dipoles within the dynamical scheme as a
random motion within the stochastic description. From the WKB
ansatz we obtain for the transition probability $P(L,T)$ over a
distance $L$ in time $T$ for a single dipole
\begin{eqnarray}
P(L,T)\propto\exp
\left[-\frac{B}{\Delta}\frac{L^{4-d}}{T^{3-d}}\right].
\label{disdip}
\end{eqnarray}
By a simple scaling argument the mean square displacement is given
by
\begin{eqnarray}
\langle\delta L^2\rangle =
T^{2H}\left[\frac{\Delta}{B}\right]^{2(1-H)}\Gamma(3-2H),
\label{msd}
\end{eqnarray}
where $\Gamma(z)$ is the Gamma function \cite{Lebedev72} and we have
introduced the Hurst exponent \cite{Feder88}
\begin{eqnarray}
H=\frac{3-d}{4-d}. \label{hurst}
\end{eqnarray}
From the scaling form in Eq. (\ref{corkpz}) and from Eq.
(\ref{disdip}) it also follows that the dynamic exponent for the
dipole sector is $z=1/H$, i.e.,
\begin{eqnarray}
z=\frac{4-d}{3-d}. \label{dyn}
\end{eqnarray}

\begin{figure}
\includegraphics[width=1.0\hsize]
{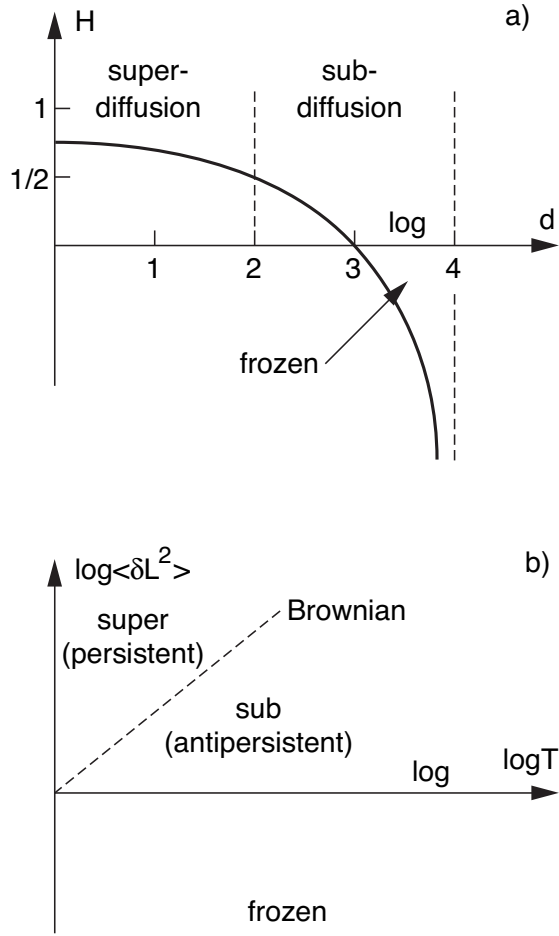}\caption{In a) we plot the Hurst exponent as function of
$d$. In b) we depict in a log-log plot the mean square deviation
$\langle\delta L^2\rangle$ as a function of $T$.} \label{fig17}
\end{figure}

In 1D we obtain $H=2/3$ and $z=3/2$ in accordance with
well-established results for the KPZ equation or, equivalently,
the noisy Burgers equation. Here the dynamic exponent for the
dipole sector agrees with the exact value. Note that the scaling
law (\ref{scal}), $z+\zeta=2$, following from Galilean invariance
and automatically implemented in the present approach, implies the
roughness exponent $\zeta = 1/2$. The Hurst exponent $H=2/3>1/2$
corresponds to a persistent random walk. Since the mean square
displacement for $H=2/3>1/2$ falls off faster than Brownian walk
($H=1/2$) we have the case of superdiffusion. In 2D, which in a
scaling context is the lower critical dimension, we have $H=1/2$,
corresponding to ordinary Brownian diffusion. The dynamic exponent
$z=2$ and according to $z+\zeta=2$ the roughness exponent
$\zeta=0$, corresponding to a smooth interface. In 3D the Hurst
exponent $H=0$ corresponding to the logarithmic case
\begin{eqnarray}
\langle\delta L^2\rangle\propto\log T~~\text{for}~~d=3.
\label{logdiff}
\end{eqnarray}
Here the mean square displacement falls off slower than Brownian
diffusion and the random motion of the dipole modes is
characterized by being antipersistent and showing subdiffusion.
For $3<d<4$ the Hurst exponent $H<0$ and the mean square
displacement decays corresponding to pinning or glassy behavior.
In 4D, which we here propose to be the upper critical dimension
for the scaling properties of the KPZ equation, the Hurst exponent
$H=-\infty$, corresponding to extreme pinning or arrested growth.
In Fig.~\ref{fig17}a we depict the Hurst exponent as a function of
dimension, in Fig.~\ref{fig17}b we plot the dipole mean square
displacement as function of time in a log-log plot.
%
\section{\label{upper} Upper critical dimension}
In addition to the scaling properties in the rough phase
characterized by a strong coupling fixed point which in 1D yields
$z=3/2$, a major open problem is the existence of an upper
critical dimension. On the basis of the singular behavior of
perturbation theory and the beta function in a Callan-Symanzik RG
scheme \cite{Wiese98}, the mapping to directed polymers
\cite{Laessig97}, and mode coupling arguments \cite{Colaiori01a},
it has been conjectured that $d=4$ is an upper critical dimension
for the KPZ equation. The behavior above 4D presumed to be complex
and maybe glassy is, however, not well understood.

Here we address the issue of an upper critical dimension within
the context of the weak noise approach and associate it with the
existence of growth modes. This is not a scaling argument but
based on the assumption that the growth mechanism and strong
coupling features of the KPZ equation depend on the existence of
propagating localized growth modes across the system. In
Sec.~\ref{field} the numerical analysis of the bound state
solution of the NLSE indicated that for $d\geq 4$ the solution is
absent. This suggest that $d=4$ within the present interpretation
plays the role of an upper critical dimension. Below we amplify
this argument by means of an application of Derrick's theorem.
\subsection{Derricks theorem}
Derrick's theorem \cite{Derrick64}, see also Refs.
\cite{Rajaraman87,Coleman85,Dodd82}, states that for a wide class
of nonlinear wave equation there does not exist stable localized
time-independent solutions with finite energy in dimensions
greater than one. The theorem effective rules out soliton-like
solutions to Lagrangian field theories in higher dimensions. Here
we sketch the simple arguments in Derrick's theorem.

Let us consider the generic Hamiltonian
\begin{eqnarray}
H=\frac{1}{2}\int d^dx~[(\nabla\phi)^2+V(\phi)+p^2], \label{dham}
\end{eqnarray}
for a scalar field $\phi$ in $d$ dimensions. $V(\phi)$ is the
potential and $p$ the momentum. From the Hamilton equations of
motion $\partial\phi/\partial t=\delta H/\delta p$ and $\partial
p/\partial t=-\delta H/\delta\phi$ we readily obtain
$\partial\phi/\partial t=p$ and $\partial p/\partial
t=\nabla^2\phi - (1/2)V'(\phi)$, or eliminating $p$, the nonlinear
wave equation
\begin{eqnarray}
\frac{\partial^2\phi}{\partial
t^2}=\nabla^2\phi-\frac{1}{2}V'(\phi).\label{dwe}
\end{eqnarray}
For a stationary field $\phi({\bf r})$ the energy is given by
\begin{eqnarray}
E=\frac{1}{2}\int d^dx~[(\nabla\phi)^2+V(\phi)], \label{de}
\end{eqnarray}
and the Euler equation
\begin{eqnarray}
\nabla^2\phi=\frac{1}{2}V'(\phi), \label{dee}
\end{eqnarray}
follows from the variational principle
\begin{eqnarray}
\frac{\delta E}{\delta\phi({\bf r})}=0. \label{var1}
\end{eqnarray}
To ensure stability we, moreover, require that the matrix
\begin{eqnarray}
\frac{\delta^2E}{\delta\phi({\bf r})\delta\phi({\bf
r}')}=[-\nabla^2+\frac{1}{2}V''(\phi)]\delta({\bf r}- {\bf
r}'),\label{var2}
\end{eqnarray}
is positive definite, i.e., $\delta^2E\geq 0$. Performing a
constrained minimization corresponding to a dilatation or scale
transformation
\begin{eqnarray}
\phi_\mu({\bf r})=\phi(\mu{\bf r}), \label{die}
\end{eqnarray}
and introducing the notation
\begin{eqnarray}
&&K=\frac{1}{2}\int d^dx (\nabla\phi)^2, \label{K}
\\
&&I=\frac{1}{2}\int d^dxV(\phi),\label{I}
\\
&&E_\mu=\frac{1}{2}\int
d^dx[(\nabla\phi_\mu)^2+V(\phi_\mu)],\label{E}
\end{eqnarray}
we obtain by substitution the relation
\begin{eqnarray}
E_\mu=\mu^{2-d}K+\mu^{-d}I.\label{e2}
\end{eqnarray}
Implementing the variational principle $\delta E/\delta\phi=0$ as
a constrained variation $dE_\mu/d\mu|_{\mu=1}=0$ we infer the
identity
\begin{eqnarray}
(2-d)K=I.\label{i1}
\end{eqnarray}
Moreover, the stability matrix $\delta^2E/\delta\phi^2\rightarrow
d^2E_\mu/d\mu^2|_{\mu=1}$ is given by
\begin{eqnarray}
\left(\frac{d^2E_\mu}{d\mu^2}\right)_{\mu=1}=(2-d)(1-d)K+d(d+1)I.
\label{var3}
\end{eqnarray}
Inserting Eq. (\ref{i1}) we finally obtain
\begin{eqnarray}
\left(\frac{d^2E_\mu}{d\mu^2}\right)_{\mu=1}=2(2-d)K.\label{var4}
\end{eqnarray}
Since $K>0$ we must require $d<2$ in order to obtain a stable
stationary solution.

This is the basic result of Derrick's theorem: The existence of
stable localized configurations $\phi({\bf r})$, i.e., with a
finite norm, as solutions to nonlinear field equations, is only
ensured below 2D. Derrick's theorem is a no-go theorem which
effectively rules out stable soliton-like solutions in higher
dimension.
%
\subsection{Nonexistence of growth modes above $d=4$}
Here we address the issue of the existence of growth modes by an
application of Derrick's theorem to the NLSE, see also Refs.
\cite{Rasmussen86,Rypdal86}. In the context of the NLSE
(\ref{chnlse}) the issue is not stability but existence of bound
states. Here the energy functional can be expressed in the form
\begin{eqnarray}
E=K+\frac{1}{2}k^2N-\frac{k_0^2}{4}I, \label{E4}
\end{eqnarray}
where $K$ is the bending energy, $N$ the norm, and $I$ the
interaction,
\begin{eqnarray}
&&K=\frac{1}{2}\int d^dx(\nabla w)^2, \label{K4}
\\
&&N=\int d^dx w^2,\label{N4}
\\
&&I=\int d^dx w^4,\label{I4}
\end{eqnarray}
and the NLSE
\begin{eqnarray}
\nabla^2w=k^2w-k_0^2 w^3, \label{NLSE}
\end{eqnarray}
follows from the variation $\delta E/\delta w=0$.

In order to establish the bound we need two identities. The first
identity is obtained by multiplying the NLSE by $w$ and
integrating over space yielding
\begin{eqnarray}
2K+k^2N-k_0^2I=0.\label{I1}
\end{eqnarray}
The second identity is obtained by constrained minimization.
Subject to the scale transformation $w({\bf r})\rightarrow
w(\mu{\bf r})$ we infer $K\rightarrow\mu^{d-2}K$,
$N\rightarrow\mu^dN$, and $I\rightarrow\mu^dI$, and applying
constrained minimization $dE/d\mu|_{\mu=1}=0$ we infer the second
identity
\begin{eqnarray}
(d-2)K+\frac{k^2}{2}dN-\frac{k_0^2}{4}dI=0.\label{I2}
\end{eqnarray}
Eliminating the bending term $K$ from the two identities we obtain
\begin{eqnarray}
k^2N=\frac{k_0^2}{4}(4-d)I. \label{I3}
\end{eqnarray}
Since $N>0$, and $I>0$ it follows that $d<4$ in order for a bound
state to exist.

This completes the proof of the nonexistence of bound states and
thus growth modes for the KPZ equation in the Cole-Hopf formulation
in dimensions larger than 4. The proof corroborates the numerical
analysis of the bound state solution.
\section{\label{sum} Summary and Conclusion}
In the present paper we have extended the weak noise approach,
previously applied in detail to the 1D noisy Burgers equation, to
the KPZ equation in higher dimensions. Three issues have been
addressed: i) kinetic pattern formation, ii) anomalous diffusion
and scaling, and iii) the upper critical dimension.

i) The weak noise WKB formulation allows a classical
interpretation of the pattern formation in the KPZ equation in the
sense that the growth of the interface is interpreted as a
dynamical deterministic network of propagating localized growth
modes. The growth modes play the role of elementary excitations
and are analogous to the vortex structures in the
Kosterlitz-Thouless theory or hedgehog structures in the
ferromagnet, see Ref. \cite{Chaikin95}. The imposed network
structure expresses the strong coupling features. Superimposed on
the network is a gas of subdominant extended diffusive modes
corresponding to the EW universality class. The dynamical
evolution of the network together with the diffusive modes defines
the kinetic pathways from an initial configuration to a final
configuration. Within the canonical weak noise scheme the network
is endowed with dynamical attributes, it carries energy, momentum,
and action. Here the action $S$ plays the particular role of a
weight function in determining the transition probability
$P\propto\exp(-S/\Delta)$, $\Delta$ is the noise strength, for a
specific kinetic pathway.

ii) The weak noise method gives access to the scaling properties
of the KPZ equation. The nonperturbative character of the WKB
approximation implies that strong coupling features might be
accessible. This is in fact the case in 1D where the exact scaling
exponents can be retrieved from the dispersion law for the growth
modes and the structure of the stationary zero-energy submanifold.
In higher D the stationary submanifold is not know and only
limited scaling results are available in the dipole sector. In the
dipole or pair mode sector, corresponding to a dilute gas of
dipole modes, the stochastic interpretation implies that the pair
modes perform random walk with Hurst exponent $H=(3-d)/(4-d)$. In
1D the interface grows stochastically with $H=2/3$, corresponding
to persistent super diffusion; in 2D, the lower critical
dimension, $H=1/2$, corresponding to ordinary Brownian motion; in
3D we have $H=0$, equivalent to logarithmic antipersistent
subdiffusion; finally, in 4D, the conjectured upper critical
dimension, $H$ diverges, corresponding to an arrested or frozen
interface. Formally, the dynamic exponent for the dipole sector is
given by $z=1/H=(4-d)/(3-d)$. In 1D we recover the well-known
result $z=3/2$; in 2D we have $z=2$ which is the weak coupling
result. In 3D the dynamic exponent diverges; in 4D we have $z=0$.
We believe that the behavior of $z$ above $d=2$ is an artifact
associated with the dipole sector; it does not reflect the true
scaling behavior of the KPZ problem.

iii) The issue of an upper critical dimension for the KPZ equation
has been much debated. Dynamical renormalization group arguments
indicates $d=4$ as the upper critical dimension; numerical
simulations, on the other hand, suggest an infinite upper critical
dimension. Here we associate the upper critical dimension with the
existence of growth modes. This is not a scaling argument but
rather associated with the idea that in the absence of a network
of localized growth modes there is no clear mechanism for the
growth of the interface. Above 4D the interface is, of course,
still governed by the nonlinear KPZ equation but we lack insight
into the actual growth mechanism if any.

This paper constitutes our present understanding of the KPZ
equation within the weak noise WKB approach. However, many open
problems remain.

i)Dynamical network: We have only analyzed the network in terms of
the asymptotic form of the growth modes. A detailed analysis of
the long time behavior requires further analysis of the field
equations in order to understand collisions and coalescence of
growth modes. Moreover, the analysis of the linear diffusive modes
and their interaction with the growth modes is incomplete. Further
analysis is required in order to determine the transition
probabilities for specific kinetic pathways.

ii)Scaling: In order to determine the scaling properties we need
to determine the stationary zero-energy manifold yielding the
stationary distribution. This is only possible in 1D; in higher D
the stationary submanifold is unknown and we only have scaling
results for the dipole sector.

iii)Dynamic renormalization group: The present approach is based
on the noise strength being the nonperturbative small parameter in
the problem. In that sense the method is not a scaling approach.
It is unclear how to relate the weak noise approach to the dynamic
renormalization group scheme.

iv) Directed polymers: An important element in our understanding
of the KPZ equation is the mapping to directed polymers in a
random medium. The relationship between the growth modes and the
wandering of polymers is an open problem.

In conclusion, the present nonperturbative weak noise scheme
represents an alternative angle of approach to the KPZ equation
and similar noise-driven problems. We emphasize that the method is
not a scaling approach based on expansions about critical
dimensions, but rather identifies the noise strength as the small
parameter in the problem. Moreover, the method is not based on
perturbation theory but on an asymptotic nonperturbative WKB or
eikonal approximation.

\begin{acknowledgments}
Discussions with A. Svane, J. Krug, and T. Halpin-Healy are
gratefully acknowledged. The present work has been supported by
the Danish Research Council.
\end{acknowledgments}

\appendix

\section{\label{app}Weak noise approach to simple random walk and the
overdamped oscillator}
\subsection{Random Walk}
\begin{figure}
\includegraphics[width=1.0\hsize]
{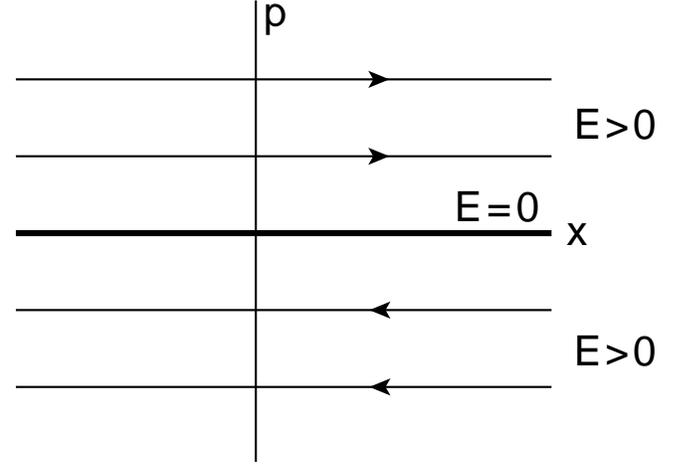}\caption{ The $(x,p)$ phase space for the random walk
case. The transient zero-energy manifold is given by $p=0$.}
\label{fig18}
\end{figure}

Simple 1D random walk is described by the Langevin equation
\begin{eqnarray}
\frac{dx}{dt}=\eta(t)~,~~~\langle\eta(t)\eta(0)\rangle=\Delta\delta(t),
\label{rw}
\end{eqnarray}
corresponding to vanishing drift $F=0$. From the weak noise scheme
we obtain the Hamiltonian
\begin{eqnarray}
H=\frac{1}{2}p^2, \label{rwham}
\end{eqnarray}
and the equations of motion
\begin{eqnarray}
\frac{dx}{dt}=p~,~~\frac{dp}{dt}=0, \label{rweq}
\end{eqnarray}
with solution from $x^i$ to $x$ in time $T$,
\begin{eqnarray}
x-x^i=p_0T~,~~p=p_0. \label{rwsol}
\end{eqnarray}
In the weak noise scheme random walk corresponds to free particle
propagation. The action is
\begin{eqnarray}
S=\frac{1}{2}p_0^2T, \label{rwaci}
\end{eqnarray}
and we infer the well-known distribution \cite{Risken89}
\begin{eqnarray}
P(x,T)\propto\exp\left[-\frac{(x-x^i)^2}{2\Delta T}\right],
\label{rwdis}
\end{eqnarray}
for the spread of random walk, i.e., the mean square displacement
$\langle(x-x^i)^2\rangle\propto\Delta T$. The transient
zero-energy manifold is given by $H=(1/2)p_0^2=0$. There is no
stationary zero-energy manifold and no saddle point in accordance
with the fact that random walk does not attain a stationary state.
In Fig.~\ref{fig18} we show the simple random walk phase space.
%
\subsection{Overdamped oscillator}
The overdamped oscillator in 1D is described by the Langevin
equation
\begin{eqnarray}
\frac{dx}{dt}=-\gamma
x+\eta(t)~,~~~\langle\eta(t)\eta(0)\rangle=\Delta\delta(t),
\label{osc}
\end{eqnarray}
corresponding to the drift $F=2\gamma x$. We obtain the
Hamiltonian
\begin{eqnarray}
H=\frac{1}{2}p^2-\gamma p x=\frac{1}{2} p(p-2\gamma x),
\label{oscham}
\end{eqnarray}
and ensuing equations of motion
\begin{eqnarray}
\frac{dx}{dt}=-\gamma x+p~,~~\frac{dp}{dt}=\gamma p, \label{osceq}
\end{eqnarray}
with solution from $x^{\text{i}}$ to $x$ in time $T$,
\begin{eqnarray}
&& x(t)=\frac{x\sinh\gamma t +
x^{\text{i}}\sinh\gamma(T-t)}{\sinh\gamma T}, \label{oscsol1}
\\
&&p(t)=\gamma e^{\gamma t}\frac{x-x^i e^{-\gamma T}}{\sinh\gamma
t}. \label{oscsol2}
\end{eqnarray}
From Eqs. (\ref{osceq}) we have $d^2x/dt^2 = \gamma^2 x$ and the
overdamped noise-driven oscillator thus corresponds to the motion
of a particle in an inverted harmonic potential.

\begin{figure}
\includegraphics[width=1.0\hsize]
{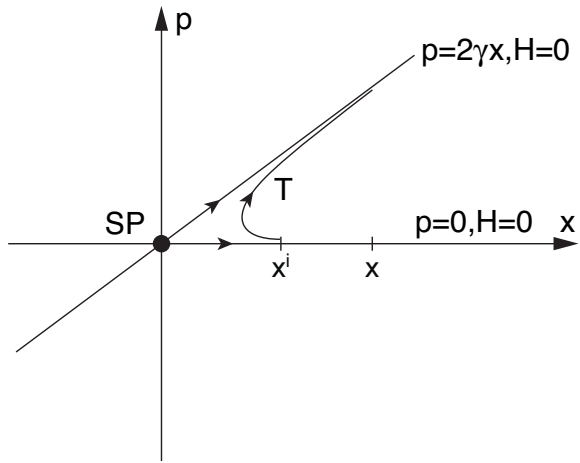}\caption{ The $(x,p)$ phase space for the overdamped
oscillator case. The transient zero-energy manifold is given by
$p=0$ and the stationary zero-energy manifold by $p=2\gamma x$.
The manifolds intersect at the saddle point (SP) $(x,p)=(0,0)$.}
\label{fig19}
\end{figure}

The zero-energy manifold $H=0$ is composed of the transient
submanifold $p=0$ and the stationary submanifold $p=2\gamma x$
intersecting at the saddle point $(x,p)=(0,0)$. From the action
$S=(1/2)\int_0^Tdt~p(t)^2$ we obtain the familiar distribution
\cite{Risken89}
\begin{eqnarray}
P(x,T)\propto\exp \left[-\frac{\gamma}{\Delta}\frac{(x-x^ie^{-\gamma
T})^2}{1-e^{-2\gamma T}}\right].~~~~ \label{oscdis}
\end{eqnarray}
For $T\rightarrow\infty$ we infer the stationary distribution
\begin{eqnarray}
P_0(x)\propto\exp\left[-\frac{\gamma x^2}{\Delta}\right].
\label{oscstat}
\end{eqnarray}
In Fig.~\ref{fig19} we have shown the phase space for the
overdamped oscillator.
We note that in both the random walk case and the overdamped
oscillator case, the weak noise method gives the exact and
familiar results; quite similar to the corresponding calculations
in the WKB approximation in quantum mechanics.

\end{document}